\documentclass[10pt,twocolumn]{iopart}
\usepackage{epsf}

\newcommand{\be}{\begin{equation}}
\newcommand{\ee}{\end{equation}}

\newcommand{\bea}{\begin{eqnarray}}

\newcommand{\eea}{\end{eqnarray}}

\newcommand{\ul}[1]{\underline{#1}}

\newcommand{\dul}[1]{\underline{\underline{#1}}}

\newcommand{\tul}[1]{\underline{\underline{\underline{#1}}}}

\newcommand{\qul}[1]{\underline{\underline{\underline{\underline{#1}}}}}

\newcommand{\tint}{\int \hspace{-3mm} \int \hspace{-3mm} \int}

\newcommand{\eps}{\epsilon}

\newcommand{\rmj}{{\rm j}}

\newcommand{\eL}{{\rm L}}

\newfont{\sfb}{cmssbx10}

\begin{document}

\title[Planar Chirality of Plasmonic Multi-Split Rings]{Planar Chirality of Plasmonic Multi-Split Rings}
\author{Luk R. Arnaut}
\address{
Time, Quantum and Electromagnetics Team, National Physical Laboratory, Hampton Road, Teddington TW11 0LW, United Kingdom \\
and\\
Department of Electrical and Electronic Engineering, Imperial College of Science, Technology and Medicine, South Kensington Campus, London SW7 2AZ, United Kingdom}
\ead{luk.arnaut@npl.co.uk}

\begin{abstract}
We develop an analytical framework for the analysis of planar chiral multi-split rings, based on a Fourier series expansion of the induced current.
The number, width, and location of each split (gap) is arbitrary. Provision is made for the possibility of inserting lumped active or passive impedance loads in the gaps.
The model demonstrates the hallmark of planar chirality and its consequent magneto-electric coupling as a function of the parameters of the splits and impedance loads. It is demonstrated that impedance loads sequenced in a clockwise or anti-clockwise fashion allow for generation of dissymetric current distributions in the plane, like for geometric (structural) chirality.
We also determine the effect of planar chirality on the relevant dyadic polarizability components of a split ring. Reflection and transmission coefficient characteristics of regular arrays of such rings are determined with the aid of the periodic method of moments.
Phase-coherent detection of the sense of handedness and incoherent detection of planar chirality {\it per se} by measurement of the field intensity are shown to be possible, in both the reflection and transmission characteristics. These results are relevant to both coherent and incoherent detection at optical and suboptical wavelengths.
\end{abstract}



(submitted 04 Nov 2008)\\

(Some figures in this article are in colour only in the electronic version)

\submitto{\JOA}

\maketitle

\section{Introduction}

Some fifteen years ago, two independent accounts of the concept of planar chirality were proposed in \cite[Ch. 8]{arna_thesis} and \cite{hech1}, seeking to generalize the notion of conventional, i.e., 3-D chirality \cite{past1,jagg1} to 2-D space. 
In fact, in \cite{arna_thesis,arnaJEWA}, 2-D chirality arose as one particular class from a reconsideration of the notions of geometrical dissymmetry \cite{past1,vell1} and polarization of electromagnetic (EM) waves in a multi-dimensional context. It lead to the suggestion that
planar chiral structures allow for the engineering of a novel class of thin-film artificial composite bianisotropic metamaterials exhibiting optical activity, which was the primary motivation for that study.

The concept of 2-D chirality is easy to grasp: when viewed from ``above'' its plane, a planar 2-D handed structure viewed looks like the optical mirror image of the same structure viewed from below the plane. This is in marked contrast with 3-D chirality, whose handedness is an intrinsic property of the structure independent of the direction of observation (wave incidence): for example, the screw sense of a given helix, which is the archetype 3-D chiral structure, is the same when viewed from either end. This demonstrates that spatial dispersion is inherent to planar chirality.

Features of planar chiral structures that are of major practical importance include their ability to be fabricated and deposited as layered thin films, their application as magnetic thin films whose properties arise from magneto-electric coupling, and their potential for generating unusually large optical activity (i.e., relatively large rotary power for the plane of polarization of an impinging wave) compared to that for 3-D chiral structures, as was numerically predicted early on \cite{arnaICEAA1995} and experimentally verified, cf., e.g., \cite{pott1}--\cite{zhan1}. 
This rotary power results from the fact that, like 3-D chiral geometries, 2-D chirality gives rise to a different sensitivity of incident waves exhibiting left- vs. right-hand circular eigenpolarizations (circular birefringence). 
The rotary power, and its extension to diffracted beams, has formed the basis for attracting wider interest for several decades, cf., e.g., \cite{fedo1}--\cite{pott2}.
 
To date, the numerical analyses of planar chiral structures have mainly been performed using full-wave numerical methods \cite{arnaICEAA1995,arnaNATO1996,arnaBIAN1997}, using a formulation of the method of moments in terms of subdomain basis function expansions for straight conducting wire segments.
In these studies, the main focus was on the calculation of optical rotation and ellipticity in the scattered field 
and on microscopic multi-polarizabilities of planar chiral structures.
Subdomain formulations provide flexibility for treating arbitrary types and shapes of planar chiral geometries. 
In this paper, instead, we use entire-domain basis functions that are computationally more efficient, more accurate, and -- although less versatile -- better suited to the particular class of planar chiral geometry of interest here, viz., asymmetrically split rings. 
Furthermore, the closed-form expressions for the induced current afforded by an entire-domain formulation offer insight into the characteristic EM features of planar chirality, and the role of the geometry parameters and impedance loading herein.

This paper is organized as follows. 
In Section \ref{sec:currdistr}, after giving a motivation for the method, we develop the Fourier expansion for the case of a single active multi-split loop with localized voltage sources across one or more of its gaps and for a passive loop illuminated by an externally incident plane wave. We show how the electromagnetic characteristics of planar chirality transpire from the expressions for the azimuthal current distribution in enantiomeric forms of the loop. The results are valid for abritrary frequencies.
In Section \ref{sec:mulitpoleexp}, we use a quasi-static approximation (for use with effective-medium theory for the homogenized medium). Again, characteristics of planar chirality are determined, now for the polarizabilities of the loop.
In Section \ref{sec:RxTx}, we turn attention to diffraction gratings of multi-split loops and determine their reflection and transmission characteristics as a function of frequency, using the periodic method of moments. The effect of planar chirality on these characteristics is demonstrated. Finally, conclusions of this theoretical study are given Section \ref{sec:concl}. 

\section{Current distribution in multi-split planar chiral ring\label{sec:currdistr}}
\subsection{Motivation}
Previous studies of planar chiral geometries have focused on straight or curved open-ended wire-type geometries, such as generalized swastikas (gammadions) \cite{arnaICEAA1995}, \cite{papa1}--\cite{taka1}, \cite{arnaBIAN1997}, planar chiral hooks (L-or J-shaped wires, or indeed most letters of the alphabet) \cite{arna_thesis,arnaJEWA,kuwa1}, scalene triangles \cite{osip1,pott0}, spirals \cite{arnaNATO1996,boru1}, etc. These works follow a ``bottom-up'' approach, in which a planar chiral geometry is synthesized via assembly of concatenated 1-D wire segments.
However, planar chirality is inherently a 2-D notion, hence it is natural to consider a canonical 2-D geometry formed by a conducting loop with unequally spaced incisions (gaps) along its circumference, producing arcs of unequal lengths. This ``top-down'' approach has the advantage of allowing for EM properties to be calculated efficiently based on the theory of loop antennas, by expanding the loop current in a Fourier series of current eigenmodes of the corresponding closed loop. As will be shown, this analytical method enables us to demonstrate in mathematical terms the essence of planar chirality. 

The connection between triple-split rings and triangles \cite{osip1} is obvious: the locations of three splits in an otherwise closed ring define the vertices of a triangle inscribed or circumscribed by the split ring. If their azimuthal locations define a scalene triangle then the split ring exhibits planar chirality; otherwise (i.e., if the associated triangle is isosceles or equilateral) the split ring exhibits one or three planes of planar symmetry, respectively. The top-down approach for studying the chiraity of split loops has a number of advantages over triangles, viz., allowing for an elegant analytic framework for the current distribution and EM properties, as well as allowing for effortless extension to multiple splits that define inscribed or circumscribed planar polygons. 

Knowledge of the distribution (mode) of the impressed or induced current in the split ring is fundamental in the calculation of electric and magnetic dipole and higher-order multipole moments and polarizabilities of planar chiral particles. The latter, in turn, form the basis for macroscopic characterization of the permittivity and permeability of the homogenized composite material as an effective medium \cite{arnaACES1997} and for its radiation, scattering, and reflection properties.
By considering each gap in turn as either a generator or a load, a multi-split loop can be conceived as a multiply-driven and multiply-loaded loop. 

\subsection{Active ring driven by localized voltage source}
A ring driven at one or more of its gaps by an impressed voltage source allows for constructing an active array whose spatial and spectral radiation characteristics are specified. For example, active surfaces combined with double-split ring arrays have been proposed for a semi-classical implementation of a spaser \cite{zhel_spaser}. It is conceivable that further integration and miniaturization of such arrays may be achieved by using an active element within the gaps of the split ring.

\subsubsection{Closed ring}
Consider an unloaded round-wire loop of radius $a$ located in free space with intrinsic impedance $\eta_0=\sqrt{\mu_0/\epsilon_0}\simeq 120\pi$ ohm.
The wire radius $r_0$ is chosen sufficiently small compared to $a$ (i.e., with King index $\kappa \stackrel{\Delta}{=}2\thinspace{\rm ln}(2\pi r_0/a)$ larger than $10$) so that the loop can be considered in its thin-wire approximation, i.e., the radial current component is negligibly small compared to the azimuthal current $I(\phi)$.
This current, when produced by a driving emf or voltage source $V$ located at azimuth $\phi=\phi_0$, i.e., $V\delta(\phi-\phi_0)$, can be expanded in a Fourier series as \cite{wu1}, \cite[Ch. 9]{king1}
\bea
I(\phi) = V y(\phi),~~~~
\eea
where
\bea
y(\phi) = - \frac{\rmj}{\pi \eta_0} \sum^{+\infty}_{n=-\infty} \frac{\exp [ -\rmj n(\phi-\phi_0)]}{a_n}
\label{eq:ydef}
\eea
is the local admittance of the loop as measured at $\phi$. In particular, the input admittance at $\phi_0=0$ is $y(0)\stackrel{\Delta}{=}Y$. A time-harmonic dependence $\exp(\rmj \omega t)$ for $V$ and $I$ is assumed and suppressed throughout. 
The Fourier coefficients of $I(\phi)$ are given by
\bea
a_n = \frac{ka}{2} A_{n+1} - \frac{n^2}{ka} A_n + \frac{ka}{2} A_{n-1},
\label{eq:def_an}
\eea
where $n=0,\pm 1, \pm 2,\ldots$ and
\bea
A_0 = \frac{1}{\pi} {\rm ln} \left ( \frac{8a}{r_0} \right )
-
\frac{1}{2} \int^{2ka}_0 \left [ \Omega_0 (x) {\rm d}x + \rmj J_0(x) \right ] {\rm d}x
\eea
\bea
A_n = A_{-n} &=& \frac{1}{\pi} \left [ K_0 \left ( \frac{n\thinspace r_0}{a} \right ) I_0 \left ( \frac{n\thinspace r_0}{a} \right ) + C_n
\right ]\nonumber\\
&~&
- \frac{1}{2} \int^{2ka}_0 \left [ \Omega_{2n}(x) + \rmj J_{2n}(x) \right ] {\rm d}x
\eea
for $n=\pm 1, \pm 2,\ldots$,
in which $I_0(\cdot)$ and $K_0(\cdot)$ are modified Bessel functions of the first and second kind, respectively, of order zero, with
\bea
C_n = \gamma + {\rm ln} \left ( 4n \right ) - 2\sum^{n-1}_{m=0} \left ( 2m+1 \right )^{-1},
\eea
in which $\gamma\stackrel{\Delta}{=} 0.577216...$ denotes the Euler--Mascheroni constant,
and where
\bea
\Omega_{n} = \frac{1}{\pi} \int^\pi_0 \sin \left ( x \sin\theta - n \theta \right ) {\rm d} \theta
\eea
and
\bea
J_{n} = \frac{1}{\pi} \int^\pi_0 \cos \left ( x \sin\theta - n \theta \right ) {\rm d} \theta
\eea
are Lommel--Weber and Bessel functions, respectively, of the first kind and order $n$. Explicit analytical approximations for $A_n$ valid for $ka < 1.3$ can be found in \cite{lo1}.

\begin{figure}[htb] \begin{center} \begin{tabular}{c} \ 
\epsfxsize=8.2cm \epsfbox{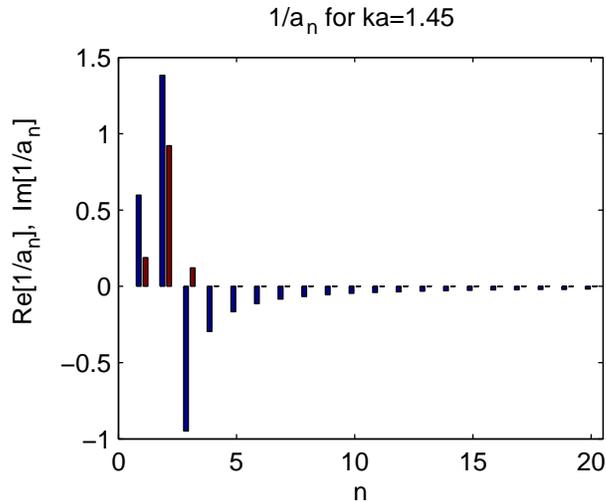}\ \\
\end{tabular} \end{center}
\caption{\label{fig:invA_ka2}
\small
Real (blue) and imaginary (red) components of $1/a_n$ ($n=0,\ldots,19$) for a closed loop with loop radius $a=20.5$ mm, wire radius $r_0=0.82$ mm and electrical size $ka=1.45$.}
\end{figure}

Figure \ref{fig:invA_ka2} shows typical values of the first twenty coefficients $1/a_n$ for a thin-wire loop at $ka=1.45$ calculated from the previous expressions based on numerical integration.

\subsubsection{Multi-split ring}
\begin{figure}[htb] \begin{center} \begin{tabular}{c} \ 
\epsfxsize=5.5cm \epsfbox{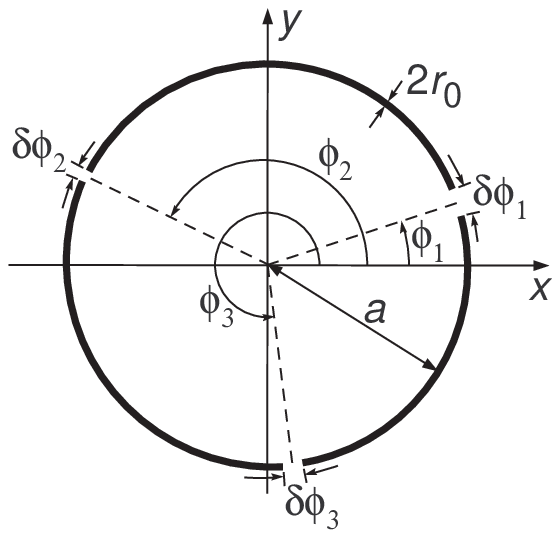}\ \\ 
(a) \\ 
\hspace{-2cm}
\epsfxsize=10.5cm \epsfbox{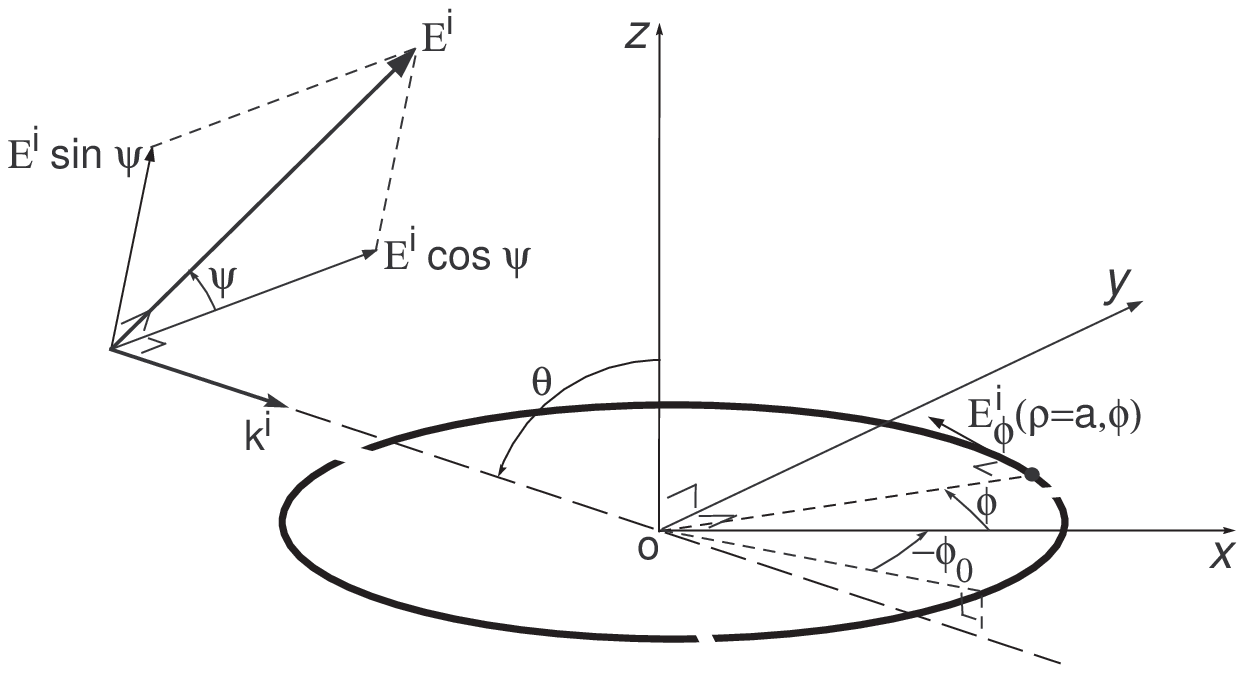}\ \\
(b)
\end{tabular} \end{center}
\caption{\label{fig:threegapring}
\small
(a) Three-split ring with loop radius $a$, wire radius $r_0$, azimuthal gap locations $\phi_k$, and gap widths $\delta \phi_k$ ($k=1,2,3$). (b) External illumination of three-split ring by plane wave. The Cartesian field components $E^{\rm i}\cos\psi$ and $E^{\rm i}\sin\psi$ are the projections of the incident electric field vector $\ul{E}^{\rm i}$ onto the $oxy$-plane of the ring and on the plane $\phi=\phi_0$, respectively. 
The angle of incident polarization $\psi$ is measured with reference to the former component.
The elevation angle of incident propagation $\theta$ is measured between the normal $oz$ to the plane of the loop and the line connecting the centre of the loop $o$ with the distant transmitter, along the direction of incidence $\ul{1}_{k^{\rm i}}=\ul{k}^{\rm i}/||\ul{k}^{\rm i}||$. 
The positive direction of the azimuthal angle of incident propagation $\phi_0$ is from the positive $ox$-axis ($\phi=0$) to the projection of the connecting line onto the $oxy$-plane.}
\end{figure}

Now consider a loop that exhibits $N$ gaps (splits) of respective permittivities $\epsilon_k$ and widths $\delta\phi_k$ that are centered at arbitrary azimuthal locations $\phi=\phi_k$ ($k=1,\ldots,N$) (Figure \ref{fig:threegapring}).
%
For simultaneous excitation of the loop at all its gaps, the total azimuthal current distribution is found by considering each ($k$th) gap in succession as a discrete voltage source with the $N-1$ remaining gaps presenting local capacitive load impedances $Z_i \delta(\phi-\phi_i) = \delta \phi_i/(\rmj \omega \epsilon_k \pi r^2_0)$ in a lumped-circuit representation of the loop for this $k$th partial excitation, with $i=1,\ldots,k-1,k+1,\ldots,N$. Inserting lumped-element impedances or more complicatd impedance circuits across the gaps allows for devising more general complex values of $Z_i$ with user-defined zeroes and poles in the complex plane as a function of frequency, including active impedances and circuits (Re$(Z) < 0$) carrying their own voltage or current source.

For the current $I_k$ generated by a delta-distributed voltage source across the $k$th gap,
\bea
I_k(\phi) = V_k y(\phi-\phi_k) - \sum^{N}_{i\not=k} Z_i I_k(\phi_i) y(\phi-\phi_i).
\label{eq:curr_Ik}
\eea
Upon summation of the $I_k(\phi_i)$ with respect to all $N-1$ gaps, i.e., summing
\bea
I_k(\phi_i) = V_k y(\phi_i-\phi_k) + \sum^{N}_{j\not=k} Z_j I_k(\phi_j) y(\phi_i-\phi_j)
\label{eq:curr_Iki}
\eea
with respect to $i$ for fixed $k$, this leads to 
\bea
&~&\sum^N_{i\not=k} \left [ 1 + Z_i \sum^N_{j\not =k} y(\phi_j-\phi_i) \right ] I_k(\phi_i) \nonumber\\
&~&~~~~= V_k \sum^N_{i\not=k} y(\phi_i-\phi_k).
\eea
Repeating the calculation for each $k$th gap source in succession, this yields a system of $N$ equations in the $N$ unknowns $I_k(\phi_i)$. Its solution can be substituted back into (\ref{eq:curr_Ik}) to yield the $k$th partial current distributions $I_k(\phi)$.
Finally, the total current associated with the simultaneously excitation of all $N$ gaps is then
\bea
I(\phi) = \sum^N_{k{=}1} I_k(\phi).
\eea

\subsection{Manifestation of planar chirality in azimuthal distribution of current}
The electromagnetic manifestation of planar chiral properties can most readily be demonstrated from the mirror symmetries of $I(\phi)$ in a two-split ring with unequal gap widths ($Z_1\not=Z_2$) or in a three-split ring with equal gaps ($Z_1=Z_2=Z_3$; {\it a fortiori} for three unequal gaps or load impedances). We refer to both cases as {\it electromagnetic} (or {\it load-induced}) {\it chirality} and {\it geometry-induced planar chirality}, respectively. Since both types are determined, respectively, by the characteristics of gaps and loop segments totalling $2\pi$, they are in a sense complementary (dual) for unloaded gaps.

For the two-split ring, the partial current distributions are
\bea
I_k(\phi) = V_k y(\phi-\phi_k) - Z_j I_k(\phi_j)y(\phi-\phi_j),
\eea
where $\{j,k\}=\{1,2\}$ with $j\not=k$.
Together with the partial gap currents $I_k(\phi_j)=V_k Y_j y(\phi_j-\phi_k)/[Y_j+y(0)]$ and $1/Z_j=Y_j$, we obtain
\bea
I(\phi) &=& V_1 y(\phi-\phi_1) + V_2 y(\phi-\phi_2) \nonumber\\
&~& - V_1 \frac{y(\phi_2-\phi_1)y(\phi-\phi_2)}{Y_2+y(0)} \nonumber\\
&~& - V_2 \frac{y(\phi_1-\phi_2)y(\phi-\phi_1)}{Y_1+y(0)}.
\label{eq:curr_tot_orig}
\eea
The current distribution of the corresponding mirror ring is obtained by a reflection in its plane with respect to an orthogonal plane at the bisecting azimuth $\phi_0=(\phi_1+\phi_2)/2$, i.e., replacing $V_1,Z_1$ by $V_2,Z_2$ or, equivalently, replacing $\phi_1,\phi_2$ by $-\phi_1,-\phi_2$, yielding
\bea
I_{\rm mirr}(\phi) 
&=& 
V_1 y(\phi+\phi_1) + V_2 y(\phi+\phi_2) \nonumber\\
&~& - V_1 \frac{y(\phi_1-\phi_2)y(\phi+\phi_2)}{Y_2+y(0)} \nonumber\\
&~& - V_2 \frac{y(\phi_2-\phi_1)y(\phi+\phi_1)}{Y_1+y(0)}\label{eq:curr_tot_mirr_a}\\
&=&
V_2 y(\phi-\phi_1) + V_1 y(\phi-\phi_2) \nonumber\\
&~& - V_2 \frac{y(\phi_2-\phi_1)y(\phi-\phi_2)}{Y_1+y(0)} \nonumber\\
&~& - V_1 \frac{y(\phi_1-\phi_2)y(\phi-\phi_1)}{Y_2+y(0)}.\label{eq:curr_tot_mirr_b}
\label{eq:curr_tot_mirr}
\eea
The expressions (\ref{eq:curr_tot_mirr_a}) and (\ref{eq:curr_tot_mirr_b}) are equivalent because $\phi_1-\phi_0=-(\phi_2-\phi_0)$.
When now replacing both $V_1,Z_1$ by $V_2,Z_2$ and $\phi_1,\phi_2$ by $-\phi_1,-\phi_2$, it is verified that $I_{\rm mirr}(\phi)$ transforms back to the original current distribution $I(\phi)$, thus demonstrating planar chirality. [Note that $y(\phi+\Delta\phi/2) \not = y(\phi-\Delta\phi/2)$ (except at $\phi=0$ and $\pi$) where $\Delta \phi \stackrel{\Delta}{=} (\phi_1- \phi_2)/2$, because $y(\cdot)$ is an even function.]
A canonical case for the two-split ring that interacts with both TE (or $s$) and TM (or $p$) polarizations arises for the special case where the splits are located at right angles ($\Delta \phi = \pm \pi/2$). In the latter case, $y(+\pi/2) = y(-\pi/2)$.

For the three-gap ring (Figure \ref{fig:threegapring}), the total current is
\bea
I(\phi)=I_1(\phi)+I_2(\phi)+I_3(\phi)
\eea 
where
\bea
I_1(\phi) &=& V_1 y(\phi-\phi_1) - Z_2 I_1 (\phi_2) y(\phi-\phi_2) \nonumber\\
&~& - Z_3 I_1(\phi_3) y(\phi-\phi_3),
\label{eq:Ionephi}
\eea
with
\bea
I_1(\phi_2) = V_1 \frac{Y_2 ~ y(\phi_2-\phi_1)}{Y_2 + y(0)} 
\frac{1 - \frac{y(\phi_2-\phi_3)}{Y_3 + y(0)}}
{1 - \frac{y(\phi_3-\phi_2)y(\phi_2-\phi_3)}{y^2(0)}}\\
\label{eq:Ionephi2}
I_1(\phi_3) = V_1 \frac{Y_3 ~ y(\phi_3-\phi_1)}{Y_3 + y(0)} 
\frac{1 - \frac{y(\phi_3-\phi_2)}{Y_2 + y(0)}}
{1 - \frac{y(\phi_2-\phi_3)y(\phi_3-\phi_2)}{y^2(0)}},
\label{eq:Ionephi3}
\eea
while expressions for $I_2(\phi)$ and $I_3(\phi)$ follow from (\ref{eq:Ionephi})--(\ref{eq:Ionephi3}) by cyclic permutation of indices.
Defining $\phi_1=0$ as the reference azimuthal orientation, it is found that for a planar achiral structure  ($\phi_2=-\phi_3$) with symmetric ($V_2=V_3$) or anti-symmetric ($V_2=-V_3$) excitation, $I(\phi)$ is a function of $y(\phi)$ and $y(\phi-\phi_2)\pm y(\phi-\phi_3)\equiv y(\phi-\phi_2)\pm y(\phi+\phi_2)$ only, viz.,
\bea
&~& \hspace{-0.5cm}
I(\phi) = 
V_1 y(\phi) + V_2 \left [ y(\phi-\phi_2) \pm y(\phi+\phi_2) \right ]\nonumber\\
&~& 
- V_1 \frac{y(\phi_2)}{Y+y(0)}
\frac{1-\frac{y(\phi_2)}{Y+y(0)}}{1-\frac{y^2(2\phi)}{y^2(0)}}
\left [ y(\phi-\phi_2)+y(\phi+\phi_2) \right ]
\nonumber\\
&~&
- V_2 \frac{y(2\phi_2)}{Y+y(0)} 
\frac{1-\frac{2y(\phi_2)}{Y+y(0)}}{1-\frac{y^2(2\phi)}{y^2(0)}}
\left [ y(\phi-\phi_2) \pm y(\phi+\phi_2) \right ]\nonumber\\
&~&
 - (V_2\pm V_2) \frac{y(\phi_2) }{Y+y(0)} \frac{1-\frac{ y(\phi_2)}{Y+y(0)}}{1-\frac{y^2(2\phi)}{y^2(0)}}
y(\phi),
\eea
in which the upper and lower signs refer to the cases $V_2=V_3$ and $V_2=-V_3$, respectively. By contrast, for a planar chiral structure, the dependence on $y(\phi-\phi_2)$ and $y(\phi-\phi_3)$ does not arise from their sum or difference.

\subsection{Passive split ring illuminated by incident plane wave\label{sec:passive}}
Passive split rings are the building blocks of metamaterials, metafilms and diffraction gratings, including high-Q and/or planar chiral spatial polarizers or filters, as well as electromagnetic bandgap (EBG) materials and disordered artificial media. Analysis of a unit-cell single inclusion is therefore of fundamental interest.

For an incident plane wave with polarization angle $\psi$ measured in the transverse plane and with elevation angle of incidence $\theta$ (Figure \ref{fig:threegapring}(b)), the normalized admittance of an unloaded closed loop is \cite[Ch. 10]{king1}
\bea
u(\phi) = \frac{\rmj}{\pi\eta_0} \sum^{+\infty}_{n=-\infty} \frac{f_n}{a_n} \exp ( -\rmj n \phi ),
\eea
in which $a_n$ is defined by (\ref{eq:def_an}) and the Fourier expansion coefficients for the incident electric field are given by
\bea
f_n &=& \frac{\rmj^n \exp(-\rmj n \phi_0)}{2} \nonumber\\
&~&\times \left \{ \sin\psi \cos\theta \left [ J_{n+1} (ka\sin\theta)+J_{n-1} (ka\sin\theta) \right ]\right. \nonumber\\
&~& \left.  + \rmj \cos \psi \left [ J_{n+1} (ka\sin\theta)-J_{n-1} (ka\sin\theta) \right ] \right \}.
\eea
In particular, for grazing incidence of a TE plane wave whose plane of incidence coincides with the plane of the loop itself ($\theta=90^\circ$, $\psi=0^\circ$), these coefficients read
\bea
f_n = \rmj^{n{-}1} n \exp (-\rmj n \phi_0) \frac{J_n(ka)}{ka},
\eea
whereas for normal incidence perpendicular to the plane of the loop ($\theta=0^\circ$, $\psi=0^\circ$), the coefficients are
\bea
f_0 = 0,~~~f_1 = \frac{\exp(-\rmj \phi_0)}{2},~~~f_{n\geq 2}=0.
\eea

For an externally illuminated one-split loop containing a single admittance load $Y_{\rm L}\equiv 1/Z_{\rm L}$ across its gap at $\phi=\phi_\eL$, the current expansion is now
\bea
I(\phi) = {\cal V} \left [ u(\phi) - \frac{u(\phi_\eL) ~y(\phi-\phi_\eL)}{Y_0 + Y_\eL}  \right ]
\label{eq:Iphi_external}
\eea
where
$
{\cal V} \stackrel{\Delta}{=} - C_a E^{\rm i}_0
$
is the emf induced by the external electric field $E^{\rm i}_0$ in the corresponding unloaded ring with circumference $C_a \stackrel{\Delta}{=}2\pi a$. This current is to be compared to
\bea
I(\phi) = V \left [ y(\phi) - \frac{y(\phi_\eL) ~y(\phi-\phi_\eL) }{Y_0 + Y_\eL} \right ]
\eea
for a loop locally driven at $\phi=0$ by a gap voltage source $V$ in the absence of external illumination.
By extension of (\ref{eq:Iphi_external}), for $N$ loads at $N$ splits located at $\phi_{j}$, the induced current is
\bea
I(\phi) = {\cal V} u(\phi) - \sum^N_{j=1} Z_j I(\phi_j) y(\phi-\phi_j).
\label{eq:Ik_ext}
\eea
With
\bea
I(\phi_{i}) = {\cal V} u(\phi_i) - \sum^N_{j=1} Z_j I (\phi_{j}) y(\phi_{i}-\phi_{j})
\label{eq:Iphi_i}
\eea
for $i=1,\ldots,N$,
the solution of the $N$ equations (\ref{eq:Iphi_i}) in the $N$ variables $I(\phi_i)$ can be expressed in matrix form as
\bea
\ul{I}_\eL = {\cal V} \left ( \dul{I} + \dul{A}_\eL \right )^{-1} \cdot \ul{u}_\eL,
\eea
where $\dul{I}$ is the unit $N\times N$ dyadic and
\bea
\ul{I}_\eL &=& [I(\phi_1) \ldots I(\phi_N)]^{\rm T}\\
\ul{u}_\eL &=& [u(\phi_1) \ldots u(\phi_N)]^{\rm T}\\
\dul{A}_\eL &=& [ y(\phi_i{-}\phi_j)\, Z_j ]_{i,j=1,\ldots,N}.
\eea
The thus obtained current components $[\ul{I}_\eL]_j \equiv I(\phi_j)$ are then substituted back into (\ref{eq:Ik_ext}) to yield $I(\phi)$.
A locally driven ring follows as a special case, by replacing $u(\phi)$ with $y(\phi)$ and omitting the self-terms $j=k$ by defining $Z_j=Z_k$ per chosen gap $j$.

In particular, for a one-split ring,
\bea
I(\phi) = {\cal V} u(\phi) - Z_\eL I(\phi_\eL)y(\phi-\phi_\eL)
\eea
or, with $I(\phi_\eL)= {\cal V} Y_\eL u(\phi_\eL)/[Y_\eL+y(0)]$ and $1/Z_j=Y_j$,
\bea
I(\phi) = {\cal V} u(\phi) - {\cal V} \frac{u(\phi_\eL)y(\phi-\phi_\eL)}{Y_\eL+y(0)}.
\label{eq:curr_tot_one}
\eea
For a two-split ring,
\bea
I(\phi) = {\cal V} u(\phi) - Z_1 I(\phi_1) y(\phi-\phi_1) - Z_2 I(\phi_2) y(\phi-\phi_2) \nonumber\\
\eea
with
\bea
I(\phi_1) &=& {\cal V} \times \nonumber\\
&~&\hspace{-1.75cm} \frac{\left [ 1+ Z_2 y(0) \right ] u(\phi_1) - Z_2 y(\phi_1-\phi_2) u(\phi_2)}{\left [ 1 + Z_1 y(0) \right ] \left [ 1 + Z_2 y(0) \right ] - Z_1 y(\phi_2-\phi_1) Z_2 y(\phi_1-\phi_2)}\nonumber\\
\\
I(\phi_2) &=& {\cal V} \times \nonumber\\
&~&\hspace{-1.75cm} \frac{\left [ 1+ Z_1 y(0) \right ] u(\phi_2) - Z_1 y(\phi_2-\phi_1) u(\phi_1)}{\left [ 1 + Z_1 y(0) \right ] \left [ 1 + Z_2 y(0) \right ] - Z_1 y(\phi_2-\phi_1) Z_2 y(\phi_1-\phi_2)}.\nonumber\\
\eea

\begin{figure}[htb] \begin{center} \begin{tabular}{c} \ 
\epsfxsize=7.5cm \epsfbox{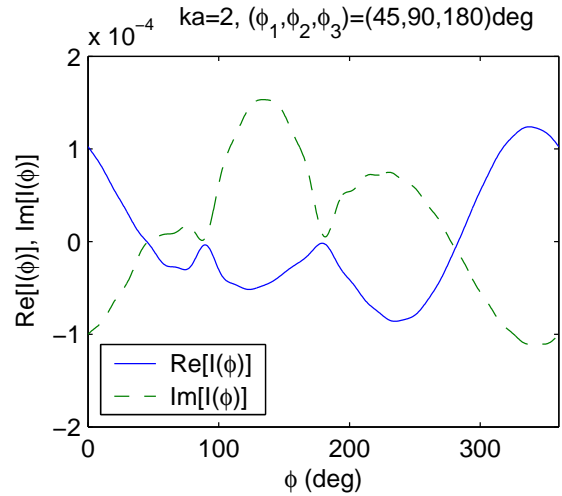}\ \\
(a) \\ 
\epsfxsize=7.5cm \epsfbox{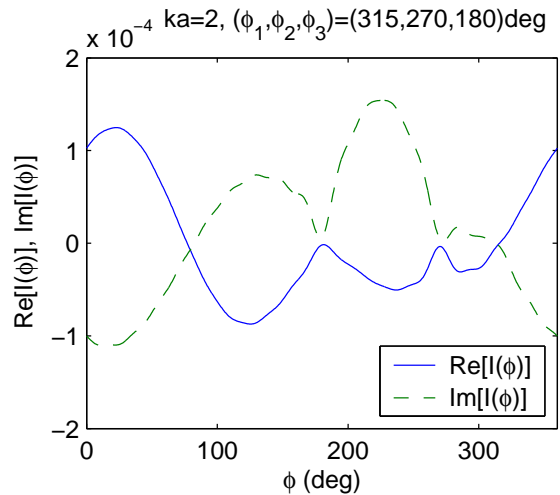}\ \\
(b) \\
\end{tabular} \end{center}
\caption{\label{fig:Iphika2_locs}\label{fig:SWIC}
\small
%
Real (blue solid curves) and imaginary (green dashed curves) parts of the current $I(\phi)$ induced in a passive three-split loop with $ka=2$, $a=20.5$ mm, $r_0=0.82$ mm, calculated from the first twenty coefficients $1/a_n$ ($n=0,\ldots,19$) for the closed loop and the coefficients $f_n$ associated with a plane wave at normal incidence and polarized along the $x$-direction, i.e., $(\theta,\phi_0,\psi)=(180^\circ,-90^\circ,0^\circ)$: (a) for a planar chiral split ring with $(\phi_1,\phi_2,\phi_3)=(45^\circ,90^\circ,180^\circ)$, $\delta \phi_1=\delta \phi_2 = \delta \phi_3 = 3^\circ$; (b) for its planar enantiomer form with $(\phi_1,\phi_2,\phi_3)=(315^\circ,270^\circ,180^\circ)$, $\delta \phi_1=\delta \phi_2 = \delta \phi_3 = 3^\circ$.}
\end{figure}

Figure \ref{fig:Iphika2_locs} shows $I(\phi)$ for a three-split ring with $a=20.5$ mm, $r_0=0.82$ mm and characterized by 
either $(\phi_1,\phi_2,\phi_3)=(45^\circ,90^\circ,180^\circ)$ or its planar chiral enantiomer 
$(\phi_1,\phi_2,\phi_3)=(315^\circ,270^\circ,180^\circ)$, with $\delta \phi_1=\delta \phi_2 = \delta \phi_3 = 3^\circ$ in both cases.
The mirror dissymmetry of the structure with respect to the $ox$-axis causes a mirror symmetry of the azimuthal current distribution with respect to $\phi=180^\circ$. Note that, at this relatively high frequency, deviations from the quasi-static result ${\rm Re}[I(\phi)]=- \rmj {\rm Im}[I(\phi)]$ occur, although these deviations are equal in magnitude for both mirror image rings.
%

\begin{figure}[htb] \begin{center} \begin{tabular}{c} \ 
\epsfxsize=7.5cm \epsfbox{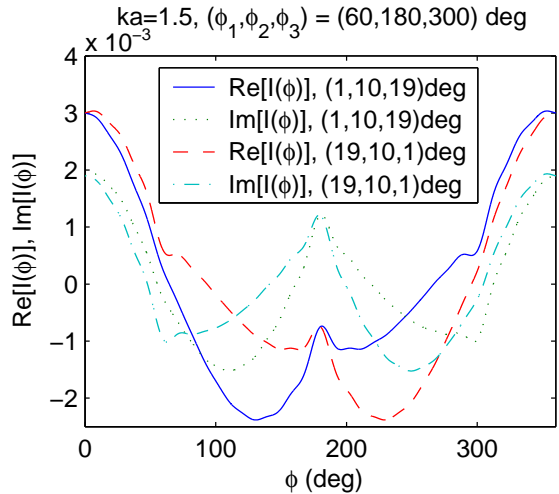}\ \\
\end{tabular} \end{center}
\caption{\label{fig:Iphika1p5_widths}
\small
Real (solid and dashed curves) and imaginary (dash-dotted and dotted curves) parts of the current $I(\phi)$ induced in a passive three-split loop with $ka=1.5$, $a=20.5$ mm, $r_0=0.82$ mm. Calculation is based on the first twenty coefficients $1/a_n$ ($n=0,\ldots,19$) for the closed loop and the coefficients $f_n$ for illumination by an obliquely incident plane wave polarized along the $y$-direction, i.e., $(\theta,\phi_0,\psi)=(30^\circ,0^\circ,0^\circ)$,
for a planar chiral split ring with
$(\phi_1,\phi_2,\phi_3)=(60^\circ,180^\circ,300^\circ)$, $(\delta \phi_1,\delta \phi_2,\delta \phi_3)= (1^\circ, 10^\circ, 19^\circ)$ and 
for its planar enantiomer with
$(\phi_1,\phi_2,\phi_3)=(60^\circ,180^\circ,300^\circ)$, $(\delta \phi_1,\delta \phi_2,\delta \phi_3)= (19^\circ, 10^\circ, 1^\circ)$.}
\end{figure}

\begin{figure}[htb] \begin{center} \begin{tabular}{c} \ 
\epsfxsize=7.5cm \epsfbox{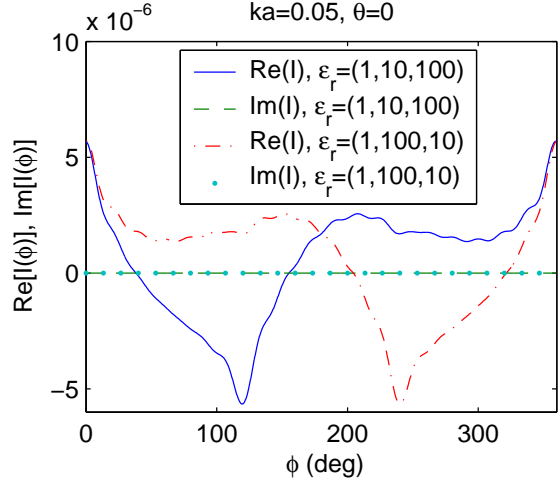}\ \\
\end{tabular} \end{center}
\caption{\label{fig:currdistr-ka0p05-epsr-theta0}
\small
Real and imaginary parts of the current $I(\phi)$ induced in a passive geometrically symmetric three-split loop with $ka=0.05$, $a=20.5$ mm, $r_0=0.82$ mm, $(\phi_1,\phi_2,\phi_3)=(0^\circ,120^\circ,240^\circ)$, $\delta \phi_1=\delta \phi_2=\delta \phi_3= 3^\circ$. Calculation is based on the first twenty coefficients $1/a_n$ ($n=0,\ldots,19$) for the closed loop and the coefficients $f_n$ for illumination by a normally incident plane wave polarized along the $y$-direction, i.e., $(\theta,\phi_0,\psi)=(0^\circ,0^\circ,0^\circ)$
 for anti-clockwise-sequenced gap permittivities 
$(\eps_{{\rm r}1},\eps_{{\rm r}2},\eps_{{\rm r}3}) = (1,10,100)$ and for the planar enantiomer (clockwise-sequenced) arrangement with 
$(\eps_{{\rm r}1},\eps_{{\rm r}2},\eps_{{\rm r}3}) = (1,100,10)$.}
\end{figure}

Alternatively, for given fixed gap locations, manifestation of planar chirality also occurs through a choice of different gap widths but whose values progress monotonically in either clockwise or counterclockwise direction. The dissymmetry of the azimuthal current distribution exists for any direction of polarization and incidence, i.e., for any value of $\theta$ and $\psi$, including normal and grazing incidence as extreme cases, although it is more pronounced in the case of oblique (glazing) incidence as presented here. Figure \ref{fig:Iphika1p5_widths} illustrates this with the case of a three-split loop at $ka=1.5$ for ($\theta=30^\circ,\phi_0=0^\circ,\psi=0^\circ)$ with $(\delta\phi_1,\delta\phi_2,\delta\phi_3)=(1^\circ,10^\circ,19^\circ)$ vs. $(\delta\phi_1,\delta\phi_2,\delta\phi_3)=(19^\circ,10^\circ,1^\circ)$ with identical gap centre locations $(\phi_1,\phi_2,\phi_3)=(60^\circ,180^\circ,300^\circ)$. 
In general, dissymmetry of the current has been found to be more pronounced when the absolute difference between the three gap widths (in this case, $9^\circ$) is increased.
Similarly, loading the gaps with lumped impedances whose values are sequential in clockwise vs. anti-clockwise direction produces dissymmetric azimuthal currents, even at exceedingly low frequencies ($ka \ll 1$) where distributed effects are small. 
Figure \ref{fig:currdistr-ka0p05-epsr-theta0} shows an example where different lumped impedances $Z_{k}$ are realized by filling the gaps $\delta \phi_k$ with different dielectric inserts of relative permittivities $\eps/\epsilon_0=1$, $10$, and $100$, arranged in either clockwise or anti-clockwise fashion. Similar dissymmetric current distributions are found at higher frequencies and for oblique incidence.
%

\section{Multipole characterization\label{sec:mulitpoleexp}}
\subsection{General considerations}
The Fourier expansion method as applied to mirror multi-split loops allows for an investigation of the effect of planar chirality on the current distribution.
The induced current is fundamental to the scattering properties and --when formed into arrays or gratings-- to the reflection and transmission characteristics. This allows for microscopic multipole modelling of a single loop and calculation of macrocopic constitutive parameters for the effective medium formed by a collection of such loops.
The latter can be carried out with the aid of an appropriate mixing rule for composite materials, e.g., using mixing rules based on Maxwell Garnett, Bruggeman, or differential equation formalisms \cite{liis1}. For an example of numerical calculation of the effective permittivity, permeability, and chirality dyadics of a homogenized metamaterial consisting of planar chiral spiral inclusions using the Maxwell Garnett formalism, cf. figures 8 and 9 of \cite{arnaNATO1996}. The higher-order multipolarizabilities of each loop can be used to define higher-order macroscopic permittivity, permeability and magneto-electrical coupling polyadics \cite{arnaACES1997}, with which the {\em macroscopic\/} effects of locality and spatial dispersion can be characterized in an explicit manner.

\subsection{Multi-polarizabilities}
In general, electric and magnetic fields at $\ul{r}^\prime$ radiated or scattered by a particle of volume $V$ with current density $\ul{J}(\ul{r})$ of harmonic time dependence $\exp(\rmj \omega t)$ can be related to its multipole moments by a Taylor expansion \cite{eyge1}--\cite{arnaAEU98} of the vector potential of $\ul{J}(\ul{r})$, i.e.,
\bea
\ul{A}(\ul{r^\prime}) = \frac{\mu_0}{4\pi}\tint_V \frac{\ul{J}(\ul{r}) \exp(-\rmj {k} ||\ul{r}^\prime-\ul{r}||)}{||\ul{r}^\prime-\ul{r}||} {\rm d}V. \label{eq:vecpot}
\eea
For a thin-wire circular loop of length $L$ carrying a local current $\ul{I}[\ul{r}(r,\phi,z)]$,
\bea
\ul{J}(\ul{r}) = I(\phi) \delta(r-a) \delta(z) \ul{1}_\phi(\phi).\label{eq:currdens}
\eea
The electric and magnetic dipole moments of this loop current are \cite{arnaAEU98}
\bea
\ul{p}_{\rm e} \stackrel{\Delta}{=} (\rmj \omega)^{-1} \tint_{V} \ul{J}(\ul{r}){\rm d}V = (\rmj \omega)^{-1} \int_{L} \ul{I}(\ell){\rm d}\ell \label{eq:def_pe}
\eea
and
\bea
\ul{p}_{\rm m} \stackrel{\Delta}{=} \frac{\mu_0}{2}  \tint_{V} \ul{\varrho}(\ul{r}) {\times} \ul{J}(\ul{r})\thinspace {\rm d}V = \frac{\mu_0}{2}  \int_{L} \ul{\varrho}(\ell) {\times} \ul{I}(\ell)\thinspace {\rm d}\ell\label{eq:def_pm},
\eea
respectively, where the second equalities in (\ref{eq:def_pe}) and (\ref{eq:def_pm}) rest on the thin-wire assumption with line coordinate $\ell$. 
These expressions apply with $\ul{E}$ and $\ul{H}\equiv\ul{B}/\mu_0$ as the source fields.

Dipole moments induced by an externally incident wave can be linked to the source fields via dipolarizability and multi-polarizability polyadics, as outlined in \ref{app:multipol} \cite{arnaBIAN1997}.
For the present purposes, we may write (\ref{eq:pe_multipole})--(\ref{eq:pm_multipole}) symbolically as \cite{arnaBIAN1998}
\bea
\ul{p}_{\rm e} = \eps_0 \/ \dul{\pi}_{\rm ee}(\ul{\nabla}) \cdot \ul{E}^{\rm inc} + \sqrt{\mu_0\eps_0} \thinspace \dul{\pi}_{\rm em}(\ul{\nabla}) \cdot \ul{H}^{\rm inc} \label{eq:pe_symb_copy}\\
\ul{p}_{\rm m} = 
\sqrt{\mu_0\eps_0} \thinspace \dul{\pi}_{\rm me}(\ul{\nabla}) \cdot \ul{E}^{\rm inc} + \mu_0 \/\dul{\pi}_{\rm mm}(\ul{\nabla}) \cdot \ul{H}^{\rm inc}, \label{eq:pm_symb_copy}
\eea
where the product operators $\dul{\pi}_{k\ell}(\ul{\nabla})$ incorporate effects of spatial dispersion to any order, through their dependence on $\ul{\nabla}$ and powers thereof, interpreted as polyadic products.
The dipolarizabilities and higher-order multi-polarizability polyadics can be extracted, for example, using a recursive multi-polarizability algorithm (RMA) based on the method of counterpropagating waves \cite{arnaBIAN1997,arnaACES1997,brew1}. For its application to planar structures, the method involves some subtleties which are outlined in \ref{app:counterprop}.

For an active split ring, 
the dipole moments are not produced by external illumination with an incident field. Its multi-polarizabilities depend instead on the applied voltage and spatial gradients of the resulting electric field. For example, for the electric field $\ul{E}_k = V_k/(\delta \phi_k)\ul{1}_{\phi_k}$ generated by $V_k$ applied across the gap $\delta \phi_k$ at $\phi=\phi_k$, the magnetic dipole moment resulting from the impressed azimuthal current defines a magneto-electric dipolarizability dyadic as
\bea
\ul{p}_{\rm m} &=& 
\sqrt{\mu_0\eps_0} \thinspace \dul{\pi}_{\rm me} \cdot \ul{E}_k.
\eea

\subsection{Electromagnetic signature of planar chirality}
As emphasized in \cite{pott0}, chirality is a purely geometric notion. While it manifests itself through symmetries and dissymmetries of the azimuthal current distribution, this does not necessarily affect electromagnetic (EM) and magnetoelectric (ME) polarizabilities through pseudo-scalarity of their elements. Indeed, the integrated azimuthal current $\int_{\rm L} I(\phi){\rm d}\phi$ (which is fundamental to calculation of all mulitpole moments) is itself insensitive to planar dissymmetries.
By subtracting the bidirectional excitation from the unidirectional one, we obtain an EM signature of planar chirality of a structure, for a given direction of incidence and polarization. For normal incidence, this signature can be defined as
\bea
\ul{\pi}^{(\pm)}_{k\ell,\alpha\beta} - \frac{1}{2} \left ( \ul{\pi}^{(+)}_{k\ell,\alpha\beta} + \ul{\pi}^{(-)}_{k\ell,\alpha\beta}\right ) \equiv \pm \frac{1}{2} \left ( \ul{\pi}^{(+)}_{k\ell,\alpha\beta} - \ul{\pi}^{(-)}_{k\ell,\alpha\beta}\right ),
\nonumber\\
\label{eq:chiralsignature}
\eea
where the upper and lower signs refer to a planar chiral structure ``$(+)$'' and its enantiomer ``$(-)$'', i.e., for a planar chiral structure ``$(+)$'' and its enantiomer ``$(-)$'' or, equivalently, for propagation in $+k_\gamma$- or $-k_\gamma$-direction, respectively ($\alpha, \beta, \gamma=x,y,z$; $\gamma \not = \alpha, \beta$). Several other measures of planar chirality have been proposed in recent years, cf., e.g., \cite{arnaBIAN1997}--\cite{boru1} and references therein.

\subsection{Numerical example}
\begin{figure}[htb] \begin{center} \begin{tabular}{c} \ 
\epsfxsize=7.5cm \epsfbox{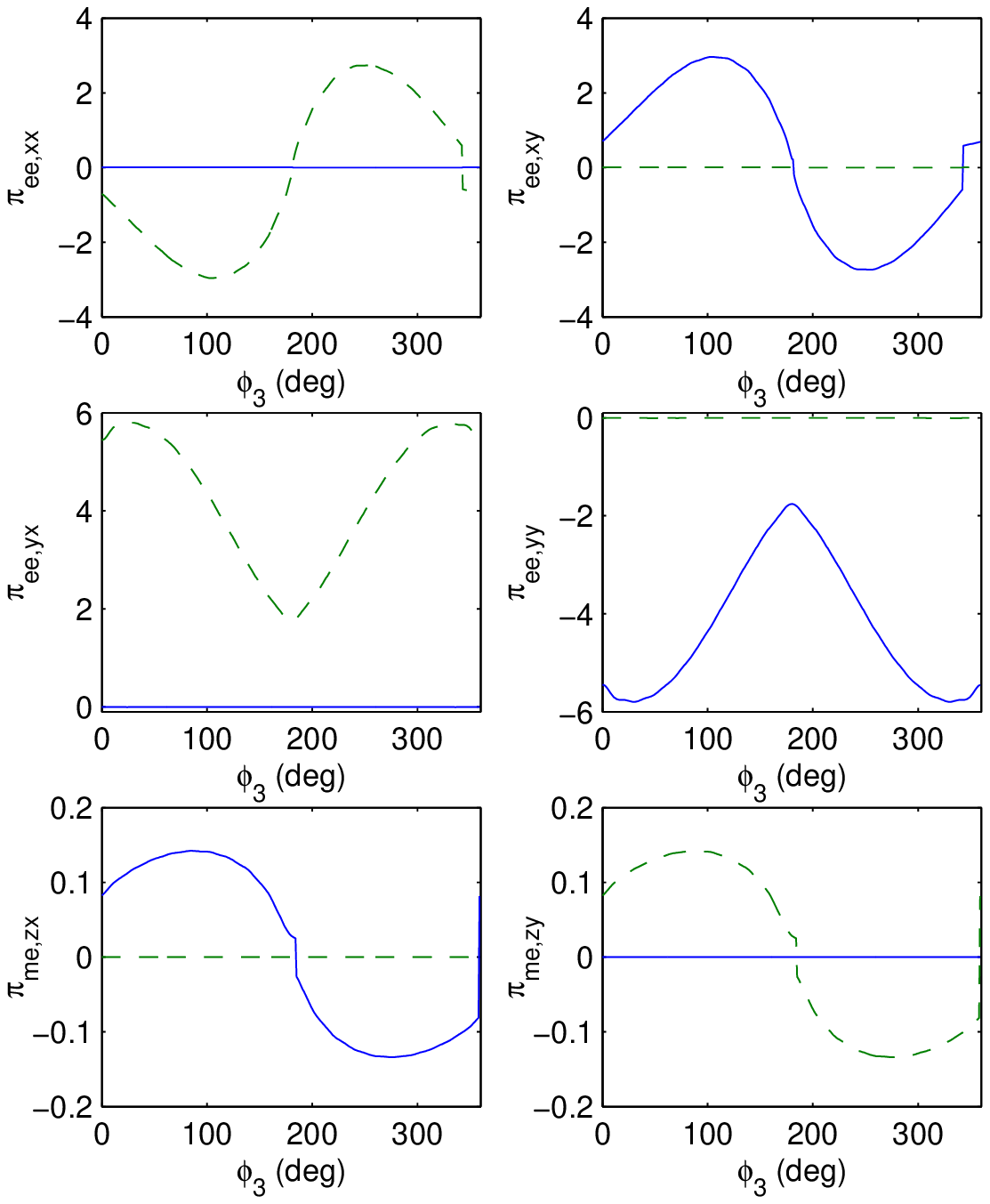}\ \\
(a) \\
 \epsfxsize=7.5cm \epsfbox{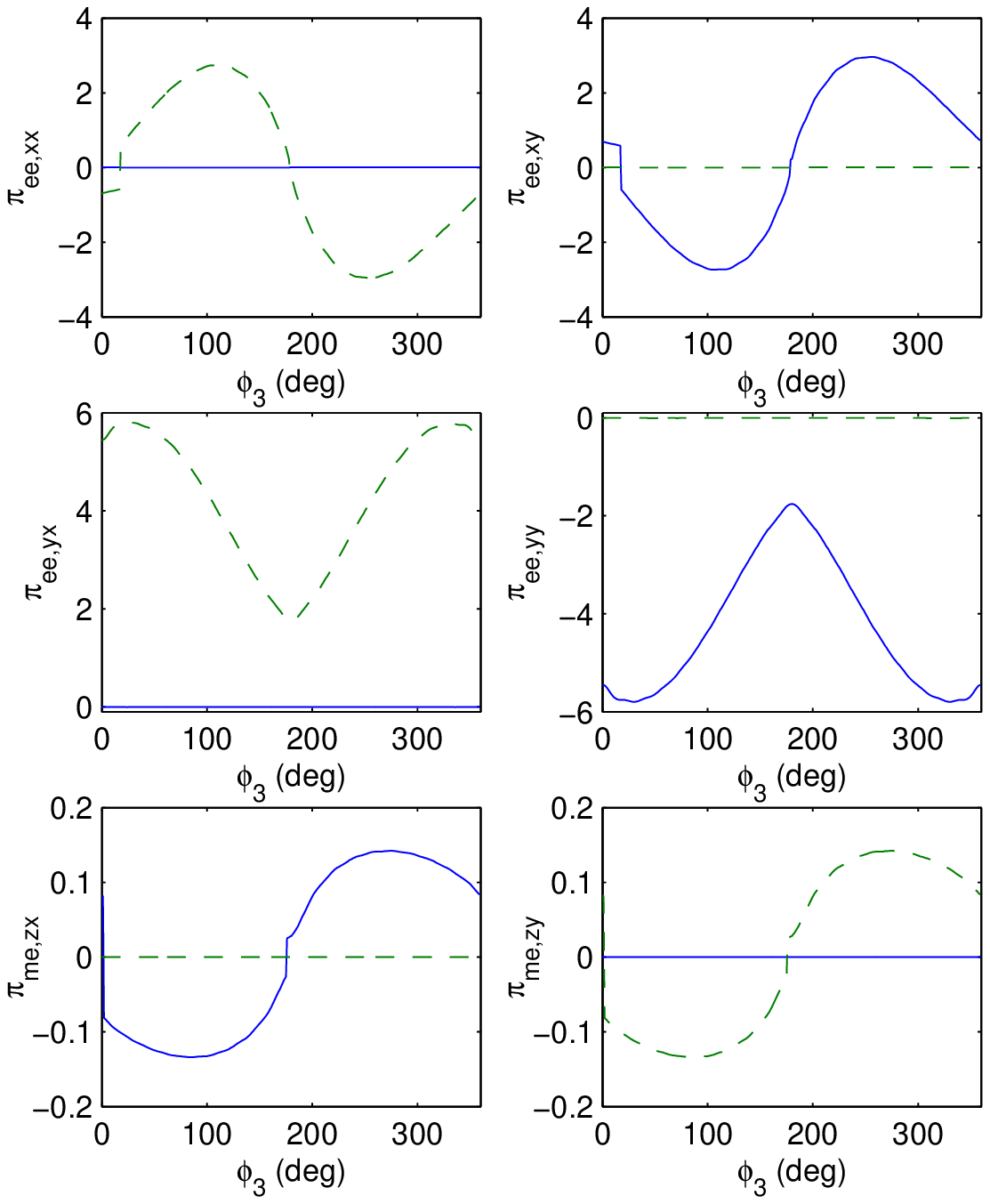}\ \\
(b) \\
\end{tabular} \end{center}
\caption{\label{fig:threesplit_ifv_splitloc_0_180deg}
\small
Real part (blue solid curve) and imaginary parts (green dashed curve) of selected polarizability elements
$\pi_{{\rm ee},\alpha\beta}$ and
$\pi_{{\rm me},z\beta}$ (in units $10^{-3}$ m$^3$)
for a single three-split ring for external plane-wave quasi-static illumination ($ka=0.05$) with
$(\phi_1,\phi_2,\phi_3)=(0^\circ,180^\circ,0^\circ\ldots360^\circ)$ and
$\delta\phi_1=\delta\phi_2=\delta\phi_3=1^\circ$:
(a) for the original structure with incidence in negative $z$-direction, and 
(b) for the enatiomeric structure (i.e., for incidence onto the original structure in positive $z$-direction). 
}
\end{figure}

Figure \ref{fig:threesplit_ifv_splitloc_0_180deg} shows elements of the electric-electric and magneto-electric polarizability dyadics 
$\pi_{\rm ee}$ and $\pi_{\rm me}$ for a three-split ring in which the locations of two 
splits are fixed at $(\phi_1,\phi_2) =(0^\circ,180^\circ)$ 
whilst the third split at $\phi_3$ is being moved in positive direction around the loop, from 
$0^\circ$ to $360^\circ$. All splits are given equal widths $\delta \phi_1=\delta \phi_2=\delta \phi_3=
1^\circ$. Quasi-static excitation of the split ring is considered here 
($ka=0.05\ll 1$; $\ul{A}(\ul{r^\prime}) \simeq [{\mu_0}/({4\pi})] \int_L {\ul{J}[\ul{r}(\ell)]}/{||\ul{r}^\prime-\ul{r}(\ell)||} {\rm d}\ell$). The series expansion for the current is limited to its first twenty Fourier coefficient terms.

From the Figure, it is observed that 
\bea
\pi_{{\rm ee},xy} &=&\pm\rmj\pi_{{\rm ee},xx},\nonumber\\ 
\pi_{{\rm ee},yx} &=&\pm\rmj\pi_{{\rm ee},yy},\nonumber\\
\pi_{{\rm me},zy} &=&\pm\rmj\pi_{{\rm me},zx},
\eea 
implying that the left-hand (LCP) and right-hand (RCP) circular wave 
polarization states $E_x \pm \rmj E_y$ are eigenpolarizations of the scattered field.
Secondly, with this choice of $\phi_1$ and $\phi_2$, two loops that have their third split at $\phi_3$ or $-\phi_3$ are enantiomers because $\phi_1$ and $\phi_2$ are diametrically opposite. Therefore, for this choice of locations for the fixed gaps, $\pi_{{\rm ee},x*}$ and $\pi_{{\rm me},z*}$ are pseudo-scalars (i.e., their signs become inverted upon taking the planar mirror image), whereas $\pi_{{\rm ee},y*}$ are true scalars. The latter follows from the fact that the two fixed gaps are both located on the $ox$-axis. Thus,
\bea
\pi^{(+)}_{{\rm ee},x*} &=& - \pi^{(-)}_{{\rm ee},x*},\nonumber\\
\pi^{(+)}_{{\rm ee},y*} &=& + \pi^{(-)}_{{\rm ee},y*},\nonumber\\
\pi^{(+)}_{{\rm me},z*} &=& - \pi^{(-)}_{{\rm me},z*},
\label{eq:relation}
\eea
where $\pi^{(+)}_{k\ell,\alpha\beta}\stackrel{\Delta}{=}\pi_{k\ell,\alpha\beta}(\phi_1,\phi_2,\phi_3)$ and $\pi^{(-)}_{k\ell,\alpha\beta}\stackrel{\Delta}{=}\pi_{k\ell,\alpha\beta}(-\phi_1,-\phi_2,-\phi_3)$ with, in this case, $\phi_1=-\phi_1$ and $\phi_2=-\phi_2$.
Equivalently, it follows that $\pi_{{\rm ee},x*}$ and $\pi_{{\rm me},z*}$ (but not $\pi_{{\rm ee},y*}$, because of the particular choice of $\phi_1$ and $\phi_2$) are signatures for planar chirality in this case and that LCP and RCP are eigenpolarizations. Obviously, the situation can be symmetrized with respect to $y$ by choosing $\phi_1=\pm 45^\circ$ and $\phi_2=\pm 225^\circ$ instead.

\begin{figure}[htb] \begin{center} \begin{tabular}{c} \ 
\epsfxsize=7.5cm \epsfbox{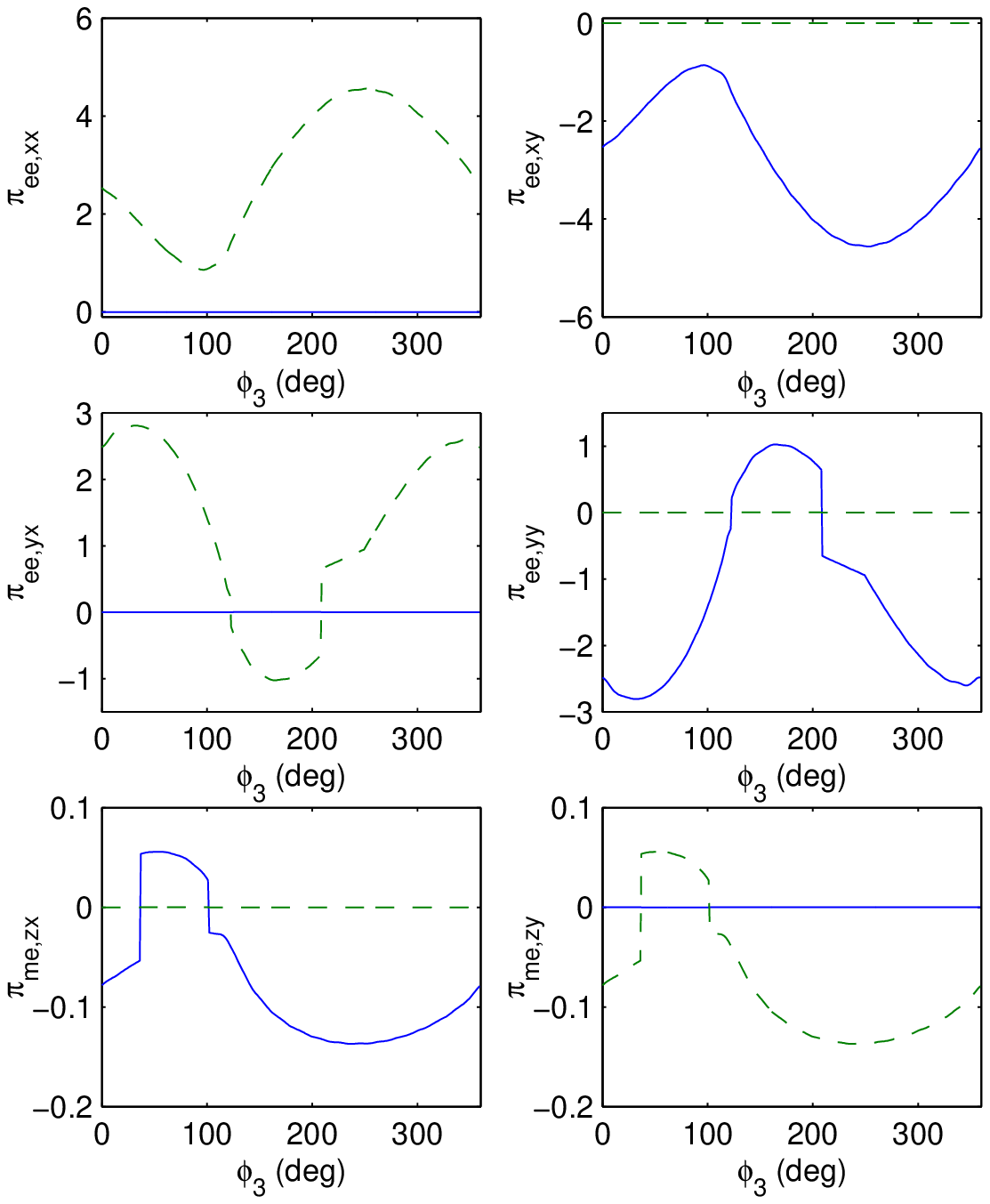}\ \\
(a) \\
 \epsfxsize=7.5cm \epsfbox{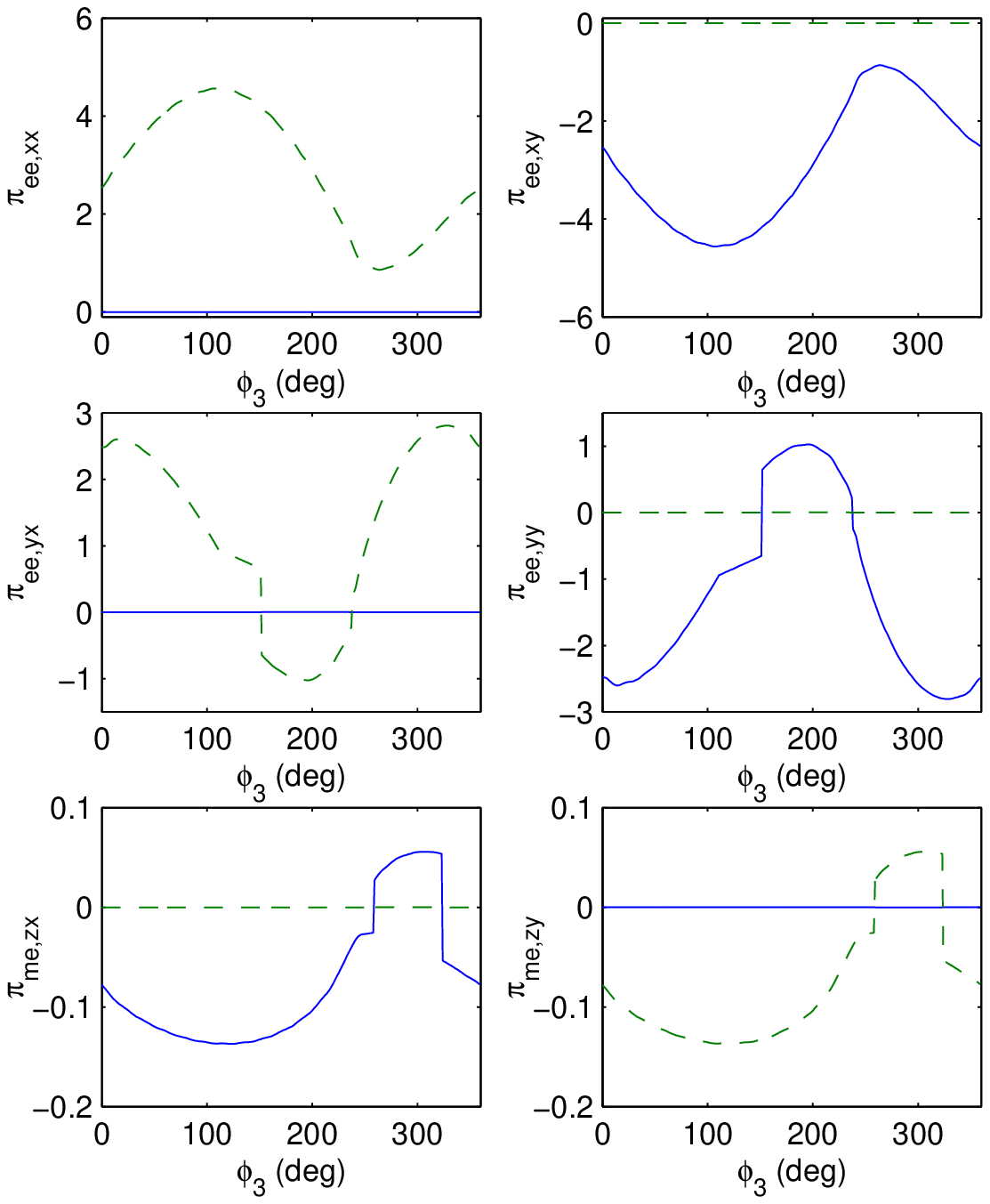}\ \\
(b) \\
\end{tabular} \end{center}
\caption{\label{fig:threesplit_ifv_splitloc_0_120deg}
\small
Same as Figure \ref{fig:threesplit_ifv_splitloc_0_180deg} but for (a) $\phi_2 = 120^\circ$ and (b) $\phi_2 = -120^\circ$.}
\end{figure}

When the two fixed splits are not diametrically opposite, the azimuthal dependence of $\pi_{k\ell,\alpha,\beta}$ is no longer symmetric, as witnessed from Figure \ref{fig:threesplit_ifv_splitloc_0_120deg} for the case $\phi_2=120^\circ$ with all other parameters left unchanged. In this case, the $\pi_{k\ell,y*}$ are no longer true scalars but also satisfy (\ref{eq:relation}). Comparison with Figure \ref{fig:threesplit_ifv_splitloc_0_180deg} further shows that the change in the $\pi_{k\ell,\alpha,\beta}$ is most pronounced when $\phi_3$ is close to $\phi_2$. The Figures also suggest that a jump caused by a current discontinuity occurs whenever one split eclipses another one, although it has been found that the size of this discontinuity becomes smaller when the $\delta(\phi_k)$ are made smaller (not shown). A more detailed investigation using a higher number of Fourier coefficients would be needed to analyze this effect in more detail.

\section{Reflection and transmission for gratings of planar chiral split rings\label{sec:PMM}\label{sec:RxTx}}
\subsection{Full-wave numerical simulation using periodic method of moments}
Large collections of multi-split loops give rise to radiated, scattered, reflected and/or transmitted waves that can be characterized with the aid of (\ref{eq:vecpot})--(\ref{eq:currdens}) provided $I(\phi)$ {\em in such collections\/} is known.
Planar arrays of closed loops and other elements have a long history of use as frequency selective surfaces (diffraction gratings) for application in reflector antennas \cite{park1,schu1}.
For such arrays, it is appropriate to use a full-wave numerical method for calculating induced currents (magnitude and spatial distribution), in order to account rigorously for mutual coupling (EM interaction) between the elements \cite{arnaNATO1996}. This mutual coupling causes the amplitudes of the Fourier expansion coefficients $u(\phi)$ and $y(\phi)$ to be different from those for a single isolated split ring.
For periodic arrays, an efficient rigorous method of analysis is the periodic method of moments (PMM) \cite{munk1}. Its implementation based on subdomain functions is used here, which has previously been detailed and validated for an array of two-split rings \cite{blac1} and other types of wire elements \cite{maly1}. 

The subdomain PMM formulation allows for easy calculation of mutual coupling between non-corresponding wire segments in different cells of the periodic array (off-diagonal elements of the impedance matrix). However, it is in principle possible to set up the entire impedance matrix for curved wire segments as well. 
Although we are focusing here only on specular reflection or transmission characteristics for normal incidence (involving neither diffraction, surface waves, nor grating lobes) at wavelengths that are large compared to the elements' size and spacing, we use a relatively large number of spectral plane-wave components $n_x\times n_y$ in the simulations for calculating the array impedance elements, viz., $n_x=n_y=10$. Previously, we have shown \cite{blac1} that small values of $n_x,n_y$ (in particular, $n_x=n_y=1$ as would be intuitively chosen in the present case) may produce highly inaccurate and unstable values for the mutual impedance elements and, hence, for the reflection or transmission characteristics.

In randomized arrays of (planar chiral) inclusions, interaction effects are implicitly taken into account within the framework of effective medium theory using mixing rules \cite{arnaNATO1996}, as mentioned before. Alternatively, the interaction between particles can be explicitly accounted for through calculation of the modified polarizability tensors \cite{arnaPIER16}, which do not require one to choose a mixing rule but instead calculates the perturbation of the microscopic current density from the particular configuration. The latter method is usable for regular and irregular arrays, at arbitrary wavelengths, but becomes cumbersome when the number of particles is large.

\subsection{Numerical results}
\begin{figure}[htb] \begin{center} \begin{tabular}{ccc} \ 
\epsfxsize=5.5cm \epsfbox{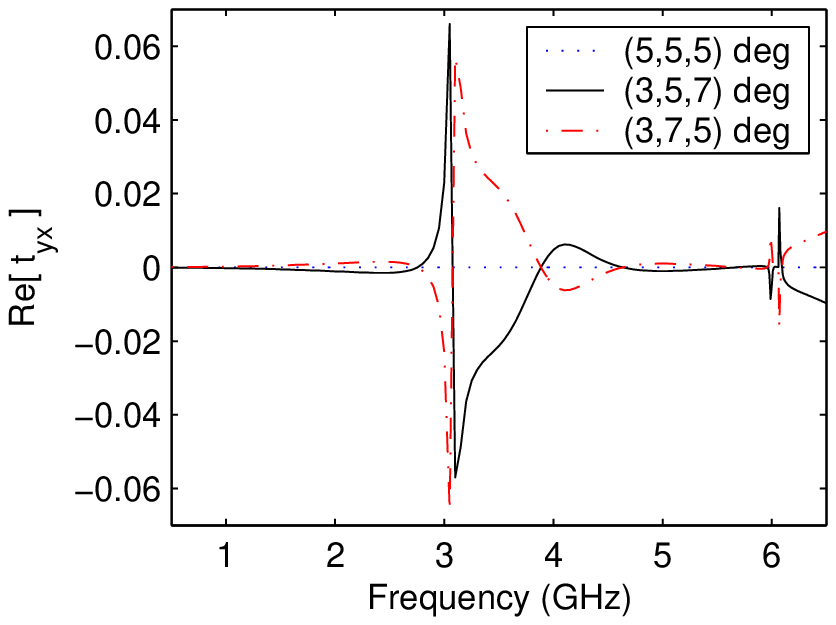}\ & \epsfxsize=5.65cm \epsfbox{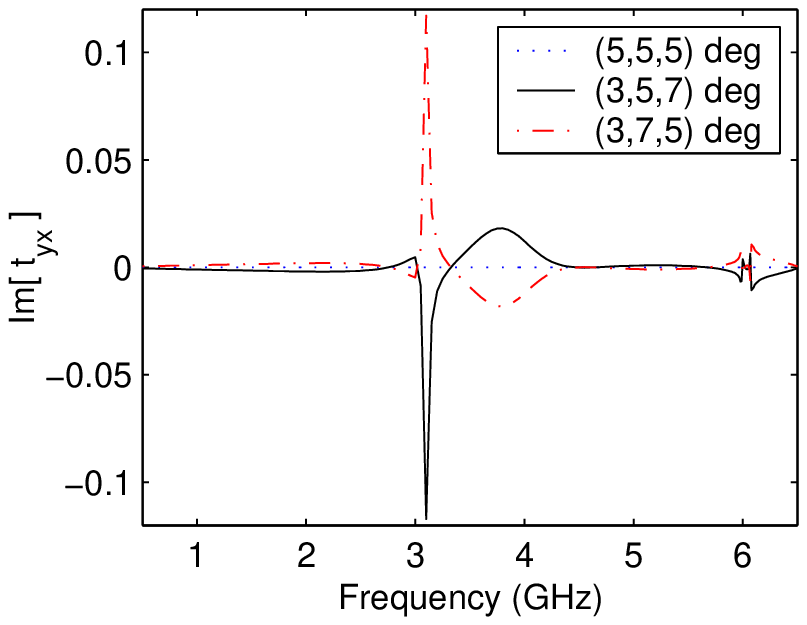}\ & \epsfxsize=5.5cm \epsfbox{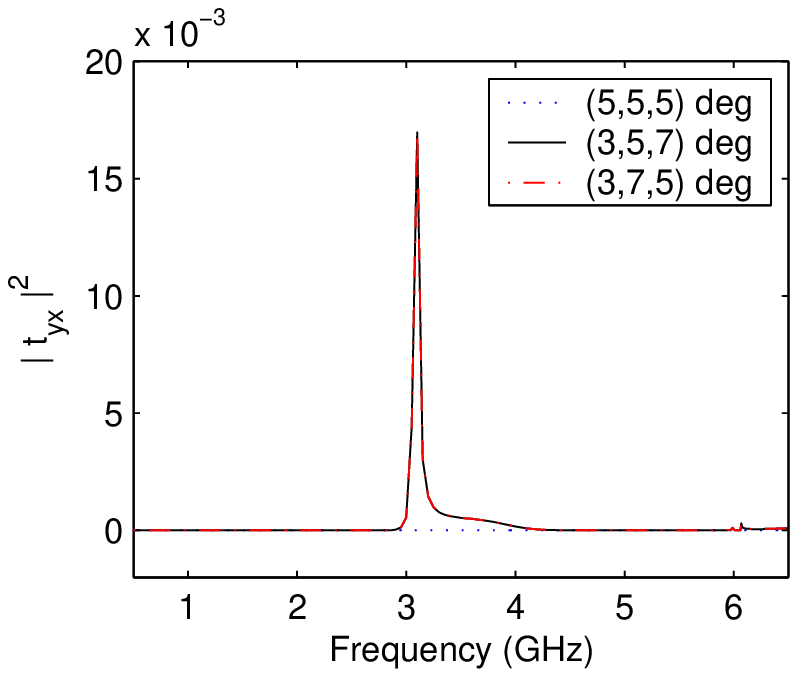}\ \\
(a) & (b) & (c)\\
\end{tabular} \end{center}
\caption{\label{fig:tcross}
\small
(a) In-phase component, (b) quadrature component, and (c) squared magnitude of cross-polarized transmission coefficient $t_{yx}$ for 
a 2-D periodic array of triple-split planar chiral loops with 
$(\phi_1,\phi_2,\phi_3)=(0^\circ,120^\circ,240^\circ)$ and $(\delta\phi_1,\delta\phi_2,\delta\phi_3)=(5^\circ,5^\circ,5^\circ)$, $(3^\circ,5^\circ,7^\circ)$, or $(3^\circ,7^\circ,5^\circ)$. Corresponding values for $t^{(\pm)}_{xy}$ are identical. }
\end{figure}

\begin{figure}[htb] \begin{center} \begin{tabular}{ccc} \ 
\epsfxsize=5.5cm \epsfbox{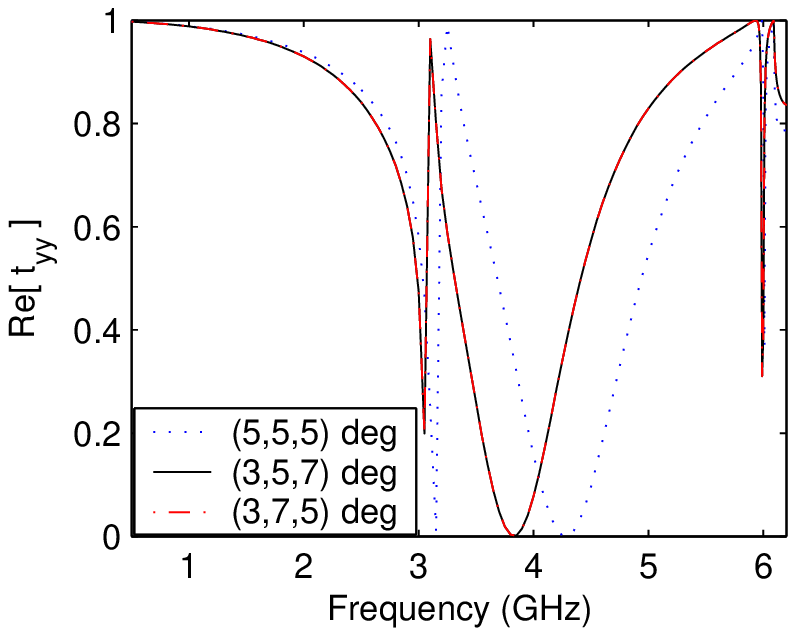}\ & \epsfxsize=5.65cm \epsfbox{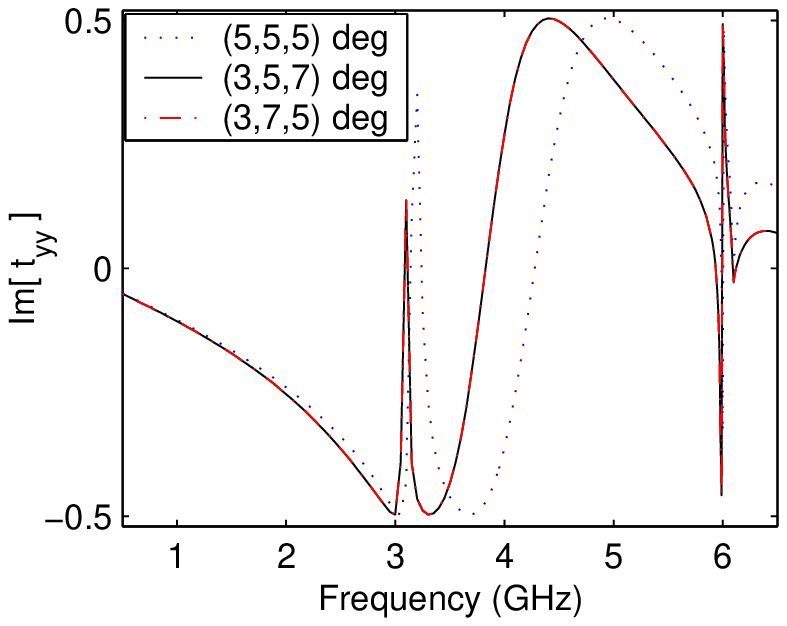}\ & \epsfxsize=5.5cm \epsfbox{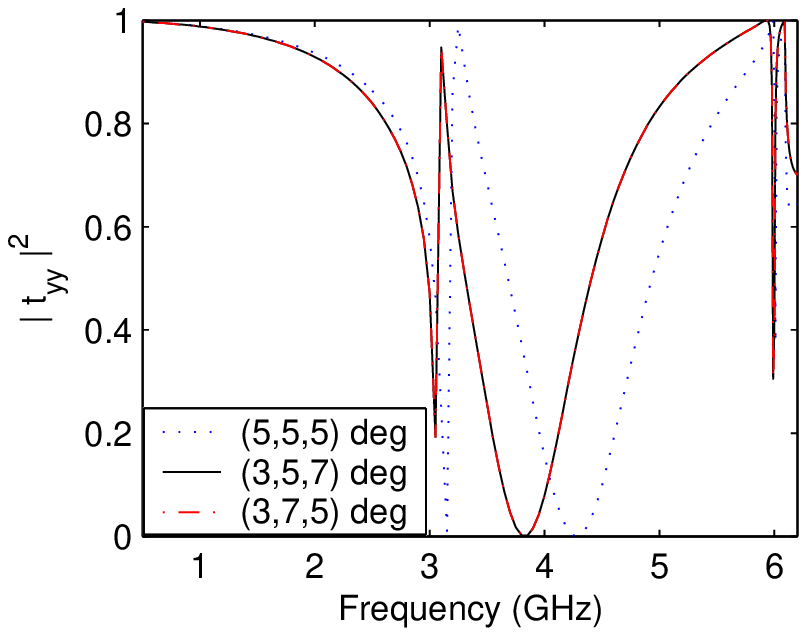}\ \\
(a) & (b) & (c)\\
\end{tabular} \end{center}
\caption{\label{fig:tco}
\small
(a) In-phase component, (b) quadrature component, and (c) squared magnitude of co-polarized transmission coefficient $t_{yy}$ for 
a 2-D periodic array of triple-split planar chiral loops with 
$(\phi_1,\phi_2,\phi_3)=(0^\circ,120^\circ,240^\circ)$ and $(\delta\phi_1,\delta\phi_2,\delta\phi_3)=(5^\circ,5^\circ,5^\circ)$, $(3^\circ,5^\circ,7^\circ)$, or $(3^\circ,7^\circ,5^\circ)$. Curves for enantiomeric forms are indistinguishable from each other.}
\end{figure}

Figures \ref{fig:tcross}(a)(b) and \ref{fig:tco}(a)(b) show frequency characteristics of the in-phase and quadrature components of the cross- and co-polarized transmission coefficients, respectively, for a periodic array of triple-split loops. The split rings form a regular lattice with period $49.2$ mm in both $x$- and $y$-directions. The loop radius and wire radius are $a=20.5$ mm and $r_0=0.82$ mm, respectively. The gap locations and split widths are listed in the Figure captions. Each of the four arcs, discretized using piecewise straight angled wire segments, was assigned four internal nodes for calculating the current distribution. The bandgap can be shifted to any desired wavelength (within the classical field model) by appropriate scaling of the geometrical parameters. The characteristic effect of the planar chirality \cite{arnaJEWA} on the phase of the coefficients, viz., an inversion of the signs of $r_{xy}$, $r_{yx}$, $t_{xy}$ and $t_{yx}$ while maintaining the signs of $r_{xx}$, $r_{yy}$, $t_{xx}$, and $t_{yy}$ is clearly witnessed. 

Figures
\ref{fig:tcross}(c) and 
\ref{fig:tco}(c) show the normalized intensity of the cross- and co-polarized 
transmission, $|t^{(\pm)}_{yx}|^2$ and $|t^{(\pm)}_{yy}|^2$,
respectively.
The cross-coupling, although small, is seen to be anti-symmetric with respect to handedness ($t^{(+)}_{yx} = - t^{(-)}_{yx}$) and is sharply peaked near the resonance frequency. The figures confirm that
the specific sense of the planar handedness is invisible to the transmitted cross-polar (non-diffracted) {\em intensity}, $|t^{(+)}_{yx}|^2$. 
Figure \ref{fig:tco} shows that the resonance frequency of the transmission coefficient for $y$- (but not for $x$-) direction exhibits a high sensitivity to planar chirality, as an indirect consequence of the different lengths of wire segments, although not to the particular sense of handedness. The effect can be used for highly-sensitive detection of asymmetries or dissymmetries with respect to specific reflection planes. Similar results were found to apply to the cross-polarized reflection coefficient ($r^{(+)}_{yx} = - r^{(-)}_{yx}$) and associated reflected intensity $|r^{(\pm)}_{yx}|^2$.

\begin{figure}[htb] \begin{center} \begin{tabular}{ccc} \ 
\epsfxsize=5.5cm \epsfbox{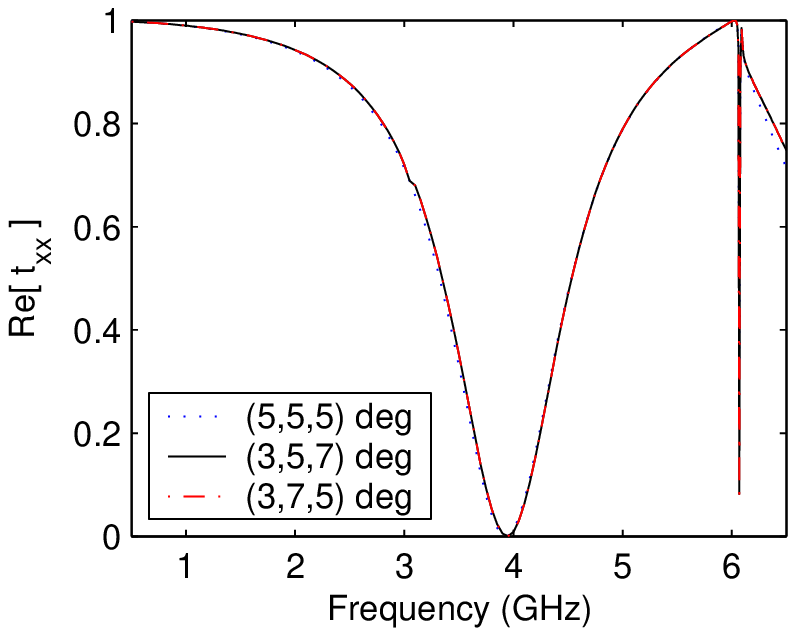}\ & \epsfxsize=5.65cm \epsfbox{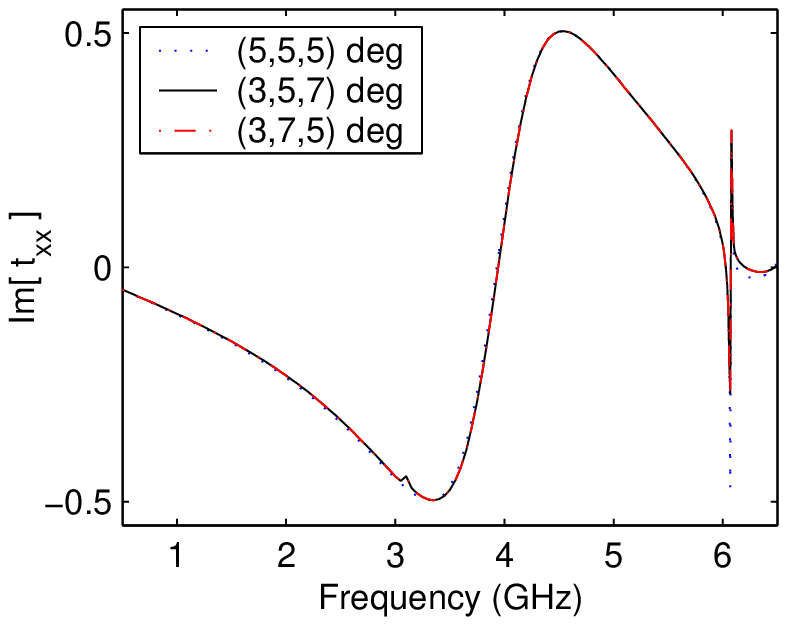}\ & \epsfxsize=5.5cm \epsfbox{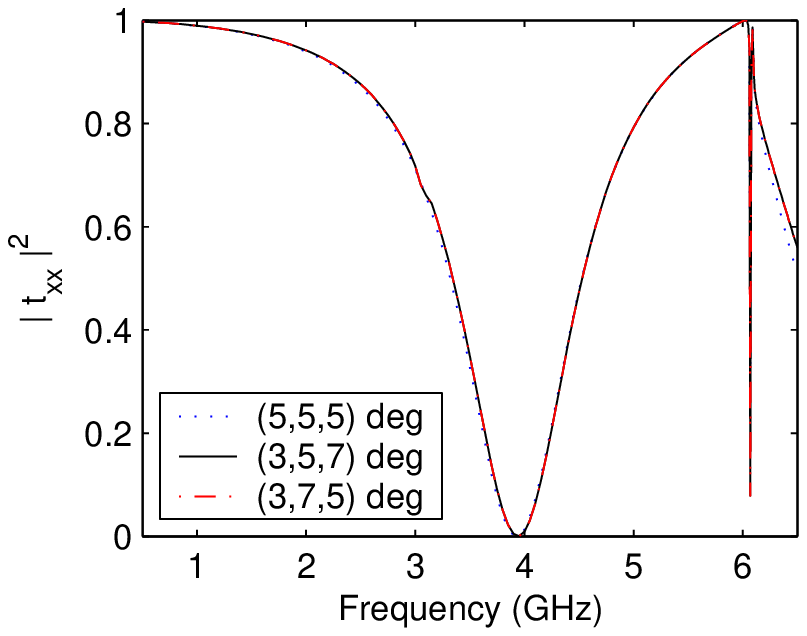}\ \\
(a) & (b) & (c)\\
\end{tabular} \end{center}
\caption{\label{fig:tcox}
\small
(a) In-phase component, (b) quadrature component, and (c) squared magnitude of co-polarized transmission coefficient $t_{xx}$ for 
a 2-D periodic array of triple-split planar chiral loops with 
$(\phi_1,\phi_2,\phi_3)=(0^\circ,120^\circ,240^\circ)$ and $(\delta\phi_1,\delta\phi_2,\delta\phi_3)=(5^\circ,5^\circ,5^\circ)$, $(3^\circ,5^\circ,7^\circ)$, or $(3^\circ,7^\circ,5^\circ)$. Curves for enantiomeric forms are indistinguishable from each other.}
\end{figure}

It should be noted that the choice of the reference azimuth $\phi=0$ for planar chiral rings carries physical significance because such rings are plano-chiral, rather than plano-semi-chiral \cite{arnaJEWA} like e.g. four-legged swastikas or other types of centro-symmetric planar structures. Because of this lack of a point centre of symmetry, circular and linear birefringence effects are coupled. The latter is apparent by comparing Figure \ref{fig:tco} with Figure \ref{fig:tcox}. Parenthetically, note that the second resonance of $t^{(\pm)}_{xx}$ exhibits a very high Q-factor. In gratings of planar chiral split rings, the linear birefringence of planar chiral rings may be suppressed by randomly orienting them in the plane, giving rise to randomization of the values of $\phi_0 \in [0,2\pi]$, cf. \cite{arnaNATO1996}. While the PMM analysis technique requires strict periodicity only in one direction, as a minimum \cite[Sec. 3.2]{munk1}, aperiodicity in the second dimension would itself introduce a preferential direction inside the plane (array effect) and hence would not eliminate the possibility of linear birefringence. Therefore, only a more cumbersome standard method of moment analysis \cite{kont1} allows for analysis of arrays of randomly rotated planar chiral rings to resolve this problem.

\section{Conclusions\label{sec:concl}}
Asymmetrically split rings serve as canonical examples of planar chiral structures, which allow for discrimination between circular polarization states of opposite handedness by a phase-coherent detector. In this paper, we investigated the effect of planar handedness on the azimuthal current distribution and polarizability coefficients of a single multi-split ring, and the reflection and transmission properties of regular 2-D arrays of such rings. It was found that geometrical as well as EM constitutive properties of the rings and their gaps allow for dissymmetric characteristics by introducing clockwise vs. anti-clockwise progressions of gap widths or permittivities of gap inserts, respectively. An analysis method based on the Fourier expansion of the current was introduced that provides an efficient alternative to full-wave simulation of an active or passive multi-split ring.

Increasing the number of gaps may be expected to give rise to more complicated frequency characteristics, because of the interplay of resonances associated with each of the wire segments whose number increases with increasing number of gaps. This requires an increasingly large number of Fourier coefficients of the wire current to be retained in order to maintain numerical accuracy.

\ack{
The author wishes to thank the reviewers for their constructive comments and suggestions.\\
}

\clearpage

\appendix

\section{Microscopic multipolarizabilities\label{app:multipol}}
The multi-polarizabilities of a point-polarizable particle can be expressed by means of a Taylor series expansion for its electric and magnetic dipole moments, as defined by (\ref{eq:def_pe}) and (\ref{eq:def_pm}), with respect to spatial derivatives of the excitation (source) fields as
\cite{arnaACES1997,arnaAEU98,buck1}
\bea
\ul{p}_{\rm e} &=& \eps_0 \dul{p}_{\rm ee} \cdot \ul{E}^{\rm inc} + \sqrt{\mu_0\eps_0} \dul{p}_{\rm em} \cdot \ul{H}^{\rm inc} \nonumber\\
&~& + \frac{1}{2!} \left [ \eps_0 \tul{q}^\prime_{\rm ee} : \ul{\nabla} \thinspace \ul{E}^{\rm inc} + \sqrt{\mu_0\eps_0} \tul{q}^\prime_{\rm em} : \ul{\nabla} \thinspace \ul{H}^{\rm inc} \right ] \nonumber\\ 
&~& + \frac{1}{3!} \left [ \eps_0 \qul{o}^{\prime\prime}_{\rm ee} \vdots~ \ul{\nabla} \thinspace \ul{\nabla} \thinspace \ul{E}^{\rm inc} + \sqrt{\mu_0\eps_0} \qul{o}^{\prime\prime}_{\rm em} \vdots~ \ul{\nabla} \thinspace \ul{\nabla} \thinspace \ul{H}^{\rm inc} \right ] \nonumber\\
&~& 
+ {\cal O} (\ul{\nabla}^3) \label{eq:pe_multipole}\\
\ul{p}_{\rm m} &=& 
\sqrt{\mu_0\eps_0} \thinspace \dul{p}_{\rm me} \cdot \ul{E}^{\rm inc} + \mu_0 \dul{p}_{\rm mm} \cdot \ul{H}^{\rm inc} \nonumber\\
&~& + \frac{1}{2!} \left [ \sqrt{\mu_0\eps_0} \tul{q}^\prime_{\rm me} : \ul{\nabla} \thinspace \ul{E}^{\rm inc} + \mu_0 \tul{q}^\prime_{\rm mm} : \ul{\nabla} \thinspace \ul{H}^{\rm inc} \right ] \nonumber\\
&~& + \frac{1}{3!} \left [ \sqrt{\mu_0\eps_0} \qul{o}^{\prime\prime}_{\rm me} \vdots~\ul{\nabla}\thinspace \ul{\nabla} \thinspace \ul{E}^{\rm inc} + \mu_0 \qul{o}^{\prime\prime}_{\rm mm} \vdots~\ul{\nabla} \thinspace \ul{\nabla} \thinspace \ul{H}^{\rm inc} \right ] \nonumber\\
&~& + {\cal O} (\ul{\nabla}^3), 
\label{eq:pm_multipole}
\eea
in which nil, single, double, triple, etc., underlined quantities represent scalars, vectors, dyadics, tryadics, etc., i.e., tensors of rank zero, one, two, three, etc., respectively.
We limit the further analysis to first-order terms in $\ul{\nabla}$.
The auxiliary tryadics $\tul{q}^\prime_{k\ell}$ are related to the actual quadrupolarizabilities $\tul{q}_{k\ell}$ via \cite{arnaAEU98}
\bea
\tul{q}^\prime_{k\ell} = \pm \tul{q}^{{\rm T}\rightarrow}_{\ell k},
\eea
where the upper and lower signs correspond to $k=\ell$ and $k\not = \ell$, respectively, and where the superscript T$\rightarrow$ denotes backward monadic transposition, i.e.,
$\left [ \tul{q}^{{\rm T}\rightarrow}_{\ell k} \right ]_{\alpha\beta\gamma} 
= \left [ \tul{q}_{\ell k} \right ]_{\gamma\alpha\beta}
$. 
From a similar expansion for the quadrupole moments $\dul{q}_{\rm e}$ and $\dul{q}_{\rm m}$,
here defined by \cite[Ch. 2]{vanb2}, \cite{arnaAEU98}
\bea
\dul{q}_{\rm e} &\stackrel{\Delta}{=}& (\rmj 2 \omega)^{-1} \int_{\rm L} [\ul{\varrho}(\ell) \ul{I}(\ell) + \ul{I}(\ell) \ul{\varrho}(\ell) ] {\rm d}\ell \\
\dul{q}_{\rm m} &\stackrel{\Delta}{=}& \frac{\mu_0}{3} \int_{\rm L} \{ \ul{\varrho}(\ell) [ \ul{\varrho}(\ell) \times \ul{I}(\ell) ] + [ \ul{\varrho}(\ell) \times \ul{I}(\ell) ] \ul{\varrho}(\ell) \} {\rm d}\ell,\nonumber\\
\label{eq:qeqm_def}
\eea
the quadrupolarizabilities $\tul{q}_{k\ell}$ are obtained from
\bea
\dul{q}_{\rm e} &=& \eps_0 \tul{q}_{\rm ee} \cdot \ul{E}^{\rm inc} + \sqrt{\mu_0\eps_0} \tul{q}_{\rm em} \cdot \ul{H}^{\rm inc} \nonumber\\
&~& + \frac{1}{2} \left [ \eps_0 \qul{o}^{\prime}_{\rm ee} : \ul{\nabla} \thinspace \ul{E}^{\rm inc} + \sqrt{\mu_0\eps_0} \qul{o}^{\prime}_{\rm em} : \ul{\nabla} \thinspace \ul{H}^{\rm inc} \right ] \nonumber\\
&~& + {\cal O} (\ul{\nabla}^2) \label{eq:qe_multipole}\\
\dul{q}_{\rm m} &=& 
\sqrt{\mu_0\eps_0} \thinspace \tul{q}_{\rm me} \cdot \ul{E}^{\rm inc} + \mu_0 \tul{q}_{\rm mm} \cdot \ul{H}^{\rm inc} \nonumber\\
&~& + \frac{1}{2} \left [ \sqrt{\mu_0\eps_0} \qul{o}^{\prime}_{\rm me} :\ul{\nabla} \thinspace \ul{E}^{\rm inc} + \mu_0 \qul{o}^{\prime}_{\rm mm} :  \ul{\nabla} \thinspace \ul{H}^{\rm inc} \right ] \nonumber\\
&~& + {\cal O} (\ul{\nabla}^2), 
\label{eq:qm_multipole}
\eea
when $I(\ell)$ is known, where $\qul{o}^{\prime(\prime)}_{k\ell} = \pm \qul{o}^{(\prime)^{{\rm T}\rightarrow}}_{\ell k}
$. By extension, the octupolarizabilities $\qul{o}_{k\ell}$ follow from a similar expansion.

For a density of $N$ particles per unit volume, macroscopic multipole moments and multi-polarizability polyadics, denoted by corresponding upper-case symbols, can be defined through $\ul{P}_{\rm e} = N \ul{p}_{\rm e}=\eps_0 \dul{P}_{\rm ee} \cdot \ul{E}^{\rm i} + \sqrt{\mu_0\eps_0} \dul{P}_{\rm em} \cdot \ul{H}^{\rm i}$, $\dul{Q}^{(\prime)}_{\rm e} = N \dul{q}^{(\prime)}_{\rm e}$, etc., in which the original external incident plane-wave fields $\ul{E}^{\rm inc}$ and $\ul{H}^{\rm inc}$ can be replaced by the local internal fields $\ul{E}^{\rm i}$ and $\ul{H}^{\rm i}$ to account for mutual interaction, with which they coincide in the dilute limit. 
The microscopic multipole expansion of $\ul{p}_{\rm e}$ and $\ul{p}_{\rm m}$ can be extended to a macroscopic multi-polarizability expansion for $\ul{P}_{\rm e}$ and $\ul{P}_{\rm m}$ and, hence, for $\dul{\epsilon}$ and $\dul{\mu}$ \cite{arnaACES1997}. 
This allows for identifying, explicitly, the contribution of spatial dispersion in the permittivity and permeability dyadics.

\section{Method of counterpropagating waves for extracting multi-polari\-zabilities of planar structures \label{app:counterprop}}
The method of counterpropagating waves \cite{brew1} can be used for the extraction of dispersion characteristics of all $36$ elements of the four $3\times 3$ electro-electric, magneto-magnetic, electro-magnetic and magneto-electric constitutive dyadics. The method has been applied to previously to 3-D chiral inclusions (helix \cite{arnaACES1997}, composite twisted hooks \cite{arnaBIAN1998}, etc.).
For analysis of 2-D planar structures, its application requires due care: waves that are incident from opposite sides experience enantiomeric forms of such structures. Therefore, the local negative superposition of waves and dipole moments is, in general, not valid in this case. 
However, for normal incidence, the magnetic field is parallel to the plane of a loop and hence will not induce currents. A similar situation occurs for grazing incidence when the electric field is perpendicular to this plane. 
For example, the excitation $(E_x,H_y,+k_z)$ characterized by $(\theta=180^\circ,\phi_0=-90^\circ,\psi=0^\circ)$ yields two of the electric-electric dyadic elements: from $p_{{\rm e},x} = \eps_0 \pi_{{\rm ee},xx} E_x(+k_z)$ and (\ref{eq:def_pe}), it follows that
\bea
\pi^{(+)}_{{\rm ee},xx}(\nabla_z) &=& \frac{ \int_{\rm L} \ul{I}(\ell) {\rm d}\ell \cdot \ul{1}_x}{{\rm j}\omega\eps_0 E_x(+k_z)} \nonumber\\
&=& - \frac{ \int^{2\pi}_0 I(\phi) \sin(\phi) {\rm d}\phi}{{\rm j}\omega\eps_0 E_x(+k_z)} \label{eq:pieexxnablaz_plus}\\
\pi^{(+)}_{{\rm ee},yx}(\nabla_z) &=& \frac{ \int_{\rm L} \ul{I}(\ell) {\rm d}\ell\cdot\ul{1}_y}{{\rm j}\omega\eps_0 E_x(+k_z)} \nonumber\\
&=& \frac{ \int^{2\pi}_0 I(\phi) \cos(\phi) {\rm d}\phi}{{\rm j}\omega\eps_0 E_x(+k_z)}, 
\label{eq:pieeyxnablaz_plus}
\eea
where $\ul{I}(\phi)=I(\phi)\ul{1}_\phi(\phi)$ is the azimuthal current flowing in the wire filament at $\phi$, whereas for propagation in $y$-direction, using the excitations
$(E_x,-H_z,+k_y)$ and $(E_x,H_z,-k_y)$ characterized by $(\theta=90^\circ,\phi_0=90^\circ,\psi=180^\circ)$ and $(\theta=90^\circ,\phi_0=-90^\circ,\psi=0^\circ)$, respectively,
\bea
\pi^{(+)}_{{\rm ee},xx}(\nabla_y) &=& \frac{1}{2} \left ( \frac{\int_{\rm L} \ul{I}(\ell) {\rm d}\ell \cdot \ul{1}_x}{{\rm j}\omega\eps_0 E_x(+k_y)}
+
\frac{\int_{\rm L} \ul{I}(\ell) {\rm d}\ell \cdot \ul{1}_x}{{\rm j}\omega\eps_0 
E_x(-k_y)}
\right )\nonumber\\\label{eq:pieexxnablay_plus}\\
\pi^{(+)}_{{\rm ee},yx} (\nabla_y) &=& \frac{1}{2} \left ( \frac{\int_{\rm L} \ul{I}(\ell) {\rm d}\ell \cdot \ul{1}_y}{{\rm j}\omega\eps_0 E_x(+k_y)} +
\frac{\int_{\rm L} \ul{I}(\ell) {\rm d}\ell \cdot \ul{1}_y}{{\rm j}\omega\eps_0
E_x(-k_y)}
\right ).\nonumber\\
\label{eq:pieeyxnablay_plus}
\eea
Likewise, from $p_{{\rm m},z} = \sqrt{\mu_0 \eps_0} \pi_{{\rm me},zx} E_x(+k_z)$ and (\ref{eq:def_pm}), or from $p_{{\rm e},x} = \sqrt{\mu_0 \eps_0} \pi_{{\rm em},xy} H_y(+k_z)$ and (\ref{eq:def_pe}), we obtain
\bea
\pi^{(+)}_{{\rm me},zx} (\nabla_z) &=& \sqrt{\frac{\mu_0}{\eps_0}} \frac{ \int_{\rm L} \ul{\varrho}(\ell) {\times} \ul{I}(\ell) \thinspace {\rm d}\ell \cdot \ul{1}_z}{2~E_x(+k_z)} \nonumber\\
&=& \sqrt{\frac{\mu_0}{\eps_0}} \frac{a \int^{2\pi}_0 I(\phi) {\rm d}\phi}{2 \thinspace E_x(+k_z)}\label{eq:pimezxnablaz_plus}\\
\pi^{(+)}_{{\rm em},xy} (\nabla_z) &=& \frac{ \int_{\rm L} \ul{I}(\ell) \thinspace {\rm d}\ell \cdot \ul{1}_x}{\rmj \omega\sqrt{\mu_0\eps_0}~H_y(+k_z)} 
=
\frac{ \int_{\rm L} \ul{I}(\ell) \thinspace {\rm d}\ell \cdot \ul{1}_x}{\rmj \omega \eps_0~E_x(+k_z)}\nonumber\\
&=& -
\frac{ \int^{2\pi}_0 I(\phi) \sin(\phi) {\rm d}\phi}{\rmj \omega \eps_0~E_x(+k_z)}.
\label{eq:piemxynablaz_plus}
\eea

For the reversely propagating excitation $(E_x,-H_y,-k_z)$, corresponding to $(\theta=0^\circ,\phi_0=-90^\circ,\psi=0^\circ)$, the obtained $p_{{\rm ee},xx}$, $p_{{\rm ee},yx}$, and $p_{{\rm me},zx}$ are again those for the enantiomeric element, which we denote by the superscript ``$(-)$''. Thus, instead of (\ref{eq:pieexxnablaz_plus}), (\ref{eq:pieeyxnablaz_plus}), and (\ref{eq:pimezxnablaz_plus}), we now have
\bea
\pi^{(-)}_{{\rm ee},xx} (-\nabla_z) &=& \frac{ \int_{\rm L} \ul{I}(\ell) {\rm d}\ell \cdot \ul{1}_x}{{\rm j}\omega\eps_0 E_x(-k_z)}
= - \frac{ \int^{2\pi}_0 I(\phi) \sin(\phi) {\rm d}\phi}{{\rm j}\omega\eps_0 E_x(-k_z)}\label{eq:pieexxnablaz_min}\nonumber\\
\\
\pi^{(-)}_{{\rm ee},yx} (-\nabla_z) &=& \frac{ \int_{\rm L} \ul{I}(\ell) {\rm d}\ell\cdot\ul{1}_y}{{\rm j}\omega\eps_0 E_x(-k_z)} = \frac{ \int^{2\pi}_0 I(\phi) \cos(\phi) {\rm d}\phi}{{\rm j}\omega\eps_0 E_x(-k_z)}  \label{eq:pieeyxnablaz_min} \nonumber\\
\\
\pi^{(-)}_{{\rm me},zx} (-\nabla_z) &=& \sqrt{\frac{\mu_0}{\eps_0}} \frac{ \int_{\rm L} \ul{\varrho}(\ell) {\times} \ul{I}(\ell) \thinspace{\rm d}\ell \cdot \ul{1}_z}{2~E_x(-k_z)} \nonumber\\
&=& \sqrt{\frac{\mu_0}{\eps_0}} \frac{a \int^{2\pi}_0 I(\phi) {\rm d}\phi}{2 \thinspace E_x(-k_z)}. 
\label{eq:pimezxnablaz_min}
\eea

Note that the superposition of bidirectional excitations, $[E_x(+k_y)+E_x(-k_y)]/2$, produces a spatially harmonic standing wave in the direction of the counterpropagation having sinusoidally varying amplitude with maximum located at $y=0$, whereas $(E_x,H_y,+k_z)$ represents a travelling plane wave whose amplitude is constant along the $y$-direction. For quasi-static excitation ($ka \ll 1$), both excitations yield identical dipolarizabilities. For $ka \not \ll 1$, however, the results can be substantially different. 
Counterpropagation has the advantage that any contribution by $H_z$, whether significant or residual, is completely cancelled, but has the disadvantage that the effect of the standing wave (i.e., contribution of spatial dispersion via the gradient of $E_x$ in $y$-direction and the quadrupolarizability elements $q^\prime_{{\rm ee},*yx}$) must be compensated for. Specifically,
\bea
p^{(+)}_{k\ell,\alpha x} E_x &=& \left ( \pi^{(+)}_{k\ell,\alpha x} - \frac{q^{(+)\prime}_{k\ell,\alpha yx}}{2}\nabla_y \right ) E_x\\
p^{(+)}_{k\ell,\alpha y} E_y &=&
\left ( \pi^{(+)}_{k\ell,\alpha y} - \frac{q^{(+)\prime}_{k\ell,\alpha xy}}{2}\nabla_x \right ) E_y\\
p^{(+)}_{k\ell,\alpha z} E_z &=& \left [ \pi^{(+)}_{k\ell,\alpha z} - \frac{1}{4}\left ( q^{(+)\prime}_{k\ell,\alpha xz}\nabla_x + q^{(+)\prime}_{k\ell,\alpha yz}\nabla_y \right ) \right ] E_z
\nonumber\\
\eea
etc., where
$k,\ell={\rm e},{\rm m}$; $\alpha=x,y,z$; and
with the spatial derivatives given by
\bea
\frac{1}{2} \nabla_y E_x &\equiv& \frac{1}{2} \left [ \nabla_y E_x(+k_y) + \nabla_y E_x(-k_y) \right ] \nonumber\\
&=& \frac{1}{2}\frac{\partial}{\partial y} [\exp(-\rmj k_y y) +  \exp(\rmj k_y y) ] E_x (\pm k_y) \nonumber\\
&=& - k_y \sin (k_y y) E_x (\pm k_y)
\eea
etc., for $-a \leq y \leq a$.
The respective dyadics $\ul{\nabla}\thinspace \ul{E}$ and $\ul{\nabla}\thinspace \ul{H}$ are given in full by 
\bea
\frac{1}{2}\ul{\nabla}\thinspace \ul{E} 
=
\left (
\begin{array}{lll}
0 & -k_x \sin(k_x x) E_y(\pm k_x) & \mp\frac{\rmj}{2} k_x \exp(\mp \rmj k_x x) E_z (\pm k_x)\\
-k_y \sin(k_y y) E_x(\pm k_y) & 0 & \mp\frac{\rmj}{2} k_y \exp(\mp \rmj k_y y) E_z (\pm k_y)\\
\mp\frac{\rmj}{2} k_z \exp(\mp \rmj k_z z) E_x (\pm k_z) & \mp\frac{\rmj}{2} k_z \exp(\mp \rmj k_z z) E_y (\pm k_z) & 0
\end{array}
\right )\nonumber\\
\eea
\bea
\frac{1}{2} \ul{\nabla}\thinspace \ul{H} 
=
\left (
\begin{array}{lll}
0 & -k_x \sin(k_x x) H_y(\pm k_x) & -k_x \sin(k_x x) H_z(\pm k_x)\\
-k_y \sin(k_y y) H_x(\pm k_y) & 0 & -k_y \sin(k_y y) H_z(\pm k_y)\\
- k_z \sin(k_z z) H_x (\pm k_z) & - k_z \sin(k_z z) H_y (\pm k_z) & 0
\end{array}
\right ).
\eea
Through cyclic permutation, $p^{(+)}_{{\rm ee},xy}$, $p^{(+)}_{{\rm ee},yy}$, and $p^{(+)}_{{\rm me},zy}$ are obtained from excitation by $(E_y,-H_x,+k_z)$ given by $(\theta=180^\circ,\phi_0=0^\circ,\psi=0^\circ)$. For the enantiomeric form, $p^{(-)}_{{\rm ee},xy}$, $p^{(-)}_{{\rm ee},yy}$, and $p^{(-)}_{{\rm me},zy}$ follow from excitation by $(E_y,H_x,-k_z)$, i.e., $(\theta=0^\circ,\phi_0=0^\circ,\psi=0^\circ)$.

For the remaining electro-magnetic and magneto-magnetic dyadic elements, excitation by $(E_x, H_z, -k_y)$ can be superimposed with a matching excitation $(-E_x,H_z,+k_y)$, 
which correspond to ($\theta=90^\circ,\phi_0=-90^\circ,\psi=0^\circ$) and ($\theta=90^\circ,\phi_0=90^\circ,\psi=0^\circ$), respectively. Similarly, excitations produced by $(E_y, H_z, +k_x)$ and $(-E_y,H_z,-k_x)$, corresponding to ($\theta=90^\circ,\phi_0=0^\circ,\psi=0^\circ$)
and
($\theta=-90^\circ,\phi_0=180^\circ,\psi=180^\circ$), can be superimposed.
Thus, from 
$2\thinspace p_{{\rm e},x} = \sqrt{\mu_0 \eps_0} p_{{\rm em},xz} [H_z(-k_y) - H_z(+k_y)]$,
$2\thinspace p_{{\rm m},z} = \mu_0 p_{{\rm mm},zz} [ H_z(-k_y) - H_z(+k_y)]$, and
(\ref{eq:def_pe}), we obtain
\bea
\pi_{{\rm em},xz} &=& \frac{1}{2} \left ( \frac{ \int_{\rm L} \ul{I}(\ell) {\rm d}\ell\cdot\ul{1}_x}{{\rm j}\omega\sqrt{\mu_0 \eps_0}H_z(-k_y)} 
\right . \nonumber\\ &~& \left.
+ \frac{ \int_{\rm L} \ul{I}(\ell) {\rm d}\ell\cdot\ul{1}_x}{{\rm j}\omega\sqrt{\mu_0 \eps_0}H_z(+k_y)} \right ) \\
\pi_{{\rm em},yz} &=& \frac{1}{2} \left ( \frac{ \int_{\rm L} \ul{I}(\ell) {\rm d}\ell\cdot\ul{1}_y}{{\rm j}\omega\sqrt{\mu_0 \eps_0}H_z(+k_x)}
\right . \nonumber\\ &~& \left. + \frac{ \int_{\rm L} \ul{I}(\ell) {\rm d}\ell\cdot\ul{1}_y}{{\rm j}\omega\sqrt{\mu_0 \eps_0}H_z(-k_x)} \right ) \\
\pi_{{\rm mm},zz} &=& \frac{1}{2} \left ( \frac{ \int_{\rm L} \ul{\varrho}(\ell) {\times} \ul{I}(\ell) \thinspace {\rm d}\ell \cdot \ul{1}_z}{H_z(\mp k_y)}
\right . \nonumber\\ &~& \left. 
+ \frac{ \int_{\rm L} \ul{\varrho}(\ell) {\times} \ul{I}(\ell) \thinspace {\rm d}\ell\cdot \ul{1}_z }{H_z(\pm k_y)} \right ),
\eea
in which $H_z(\pm k_x)$ and $H_z(\pm k_y)$ refer to the magnetic field associated with waves propagating in diametrically opposite directions. If the field components $H_\alpha$ are obtained from the same set of counterpropagating waves as for $E_\alpha$, i.e., now through their negative superposition, then again a standing wave results. 
This standing wave is now in quadrature with respect to the one for the electric-field excitation and extends along the direction of propagation, e.g.,
\bea
&~& \frac{1}{2} \left [ \nabla_z H_y(+k_z) - \nabla_z H_y(-k_z) \right ] \nonumber\\
&~&~~~=
\frac{1}{2} \frac{\partial}{\partial z} \left [ \exp(\rmj k_z z) - \exp (-\rmj k_z z) \right ] H_y\nonumber\\
&~&~~~=
\rmj k_z \cos(k_z z) H_y,
\eea
which, for the planar structure, is to be evaluated at $z=0$.
Note that a 2-D planar element (irrespective of whether it is planar chiral, symmetric, or asymmetric) exhibits nonvanishing spatial dispersion in this plane because of the nonzero spatial extent of the element in its plane. 
Specifically,
for grazing incidence, $E_x$ (resp. $E_y$) traverses a finite length of the element during propagation in $\pm y$- (resp. $\pm x$-) direction and thus contributes to $p_{{\rm m},z}$. This does not occur for these polarizations when propagation is in $\pm z$-direction, because of the zero thickness of any 2-D element in this direction.
Similarly, $H_x$ and $H_y$ affect the induced overall $\ul{p}_{\rm e}$ for grazing incidence ($\pm k_y$ or $\pm k_x$, respectively).

The elements of the quadrupolarizability tryadics are obtained by extension: from (\ref{eq:qeqm_def}) and with the same plane-wave excitations as for the dipolarizabilities, we obtain, up to first order in $\ul{\nabla}$,
\bea
q^{(+)}_{{\rm ee},\alpha\beta x} &=& \frac{ \int_{\rm L} [\ul{\varrho}(\ell)\ul{I}(\ell) + \ul{I}(\ell)\ul{\varrho}(\ell)] {\rm d}\ell : \ul{1}_\alpha \ul{1}_\beta}{{\rm j}2\omega\eps_0 E_x(+k_z)}
\eea
\bea
&~& \hspace{-0.75cm} q^{(+)}_{{\rm me},\alpha\beta x} \nonumber\\
&~& \hspace{-0.75cm} = \sqrt{\frac{\mu_0}{\eps_0}} \frac{ \int_{\rm L} \{ \ul{\varrho}(\ell) [ \ul{\varrho}(\ell) \times \ul{I}(\ell)] + [ \ul{\varrho}(\ell) \times \ul{I}(\ell)] \ul{\varrho}(\ell) \} {\rm d}\ell : \ul{1}_\alpha\ul{1}_\beta}{3~E_x(+k_z)}  
\nonumber\\
\eea 
etc. These allow for computation of first-order corrections of the dipolarizabilities for spatial dispersion effects.


\clearpage

\section*{References}

\begin{thebibliography}{99}
{\small
\bibitem{arna_thesis} Arnaut L R Apr 1994 \it Analysis and Design of Lossy Chirals and Biisotropics \rm PhD Thesis, University of Manchester Institute of Science and Technology (UMIST), Manchester, UK
%
\bibitem{hech1} Hecht L and Barron L D Aug 1994 Rayleigh and Raman optical activity from chiral surfaces \it Chem. Phys. Lett. \rm {\bf 225}(4-6) 525--30
%
\bibitem{past1} Pasteur L 1848 \it \OE vres \rm Pt. iii {\bf 24} 457. Paris: Acad\'{e}mie des Sciences
%
\bibitem{jagg1} Jaggard D L, Mickelson A R and Papas C H Feb 1979 On electromagnetic waves in chiral media \it Appl. Phys. \rm {\bf 18}(2) 211--6
%
\bibitem{arnaJEWA} Arnaut L R Nov 1997 Chirality in multi-dimensional space with application to electromagnetic characterization of plano - chiral, plano~-~semi-chiral and axi~-~chiral media \it J. Electromagn. Waves Applic. \rm {\bf 11}(11) 1459--82
%
\bibitem{vell1} Velluz L, Legrand M and Grosjean M 1965 \it Optical Circular Dichroism \rm New York, NY: Weinheim Verlag Chemie, Academic
%
\bibitem{arnaICEAA1995} Arnaut L R and Davis L E Dispersion characteristics of planar chiral structures \it Proc. 4th Int. Conf. Electromagn. Adv. Applic. (ICEAA) \rm (12--15 Sep 1995, Torino, Italy) (Swanley, UK: Nexus Media) 381--4
%
\bibitem{fedo1} Fedorov F I 1959 On the optical activity in crystals: I. The law of conservation of energy and the optical activity tensors \it Opt. Spectrosc. \rm {\bf 6}(1) 49--53
%
\bibitem{pott1} Potts A, Papakostas A, Zheludev N I, Coles H J, Greef R and Bagnall D M May 2003 Planar chiral meta-materials for photonic devices \it J. Mater. Sci.: Mater. Electron. \rm {\bf 14}(5--7) 393--5
%
\bibitem{papa1} Papakostas A, Potts A, Bagnall D M, Prosvirnin S L, Coles H J, and Zheludev N I Mar 2003 Optical manifestations of planar chirality \it Phys. Rev. Lett. \rm {\bf 90}(10) 107404
%
\bibitem{vall1} Vallius T, Jefimovs K, Turunen J, Vahimaa P and Svirko Y Jul 2003 Optical activity in subwavelength-period arrays of chiral metallic particles \it Appl. Phys. Lett. \rm {\bf 83}(2) 234--6
%
\bibitem{bede1} Bedeaux D, Osipov M A and Vlieger J Dec 2004 Reflection of light at structured chiral interfaces \it J. Opt. Soc. Am. {\rm A} \rm {\bf 21}(12) 2431--41
%
\bibitem{canf1} Canfield B K, Kujala S, Kauranen M, Jefimovs K, Vallius T and Turunen J Apr 2005 Remarkable polarization sensitivity of gold nanoparticle arrays \it Appl. Phys. Lett. \rm {\bf 86}(18) 183109
%
\bibitem{zhan1} Zhang W, Potts A, Papakostas A and Bagnall D M May 2005 Intensity modulation and polarization rotation of visible light by dielectric planar chiral metamaterials \it Appl. Phys. Lett. \rm {\bf 86}(23) 231905
%
\bibitem{pros_PRE} Prosvirnin S L and Zheludev N I 2005 Polarization effects in the diffraction of light by a planar chiral structure \it Phys. Rev. {\rm E} \rm {\bf 71}(3) 037603
%
\bibitem{kuwa1} Kuwata-Gonokami M, Saito N, Ino Y, Kauranen M, Jefimovs K, Vallius T, Turunen J, and Svirko Y Nov 2005 Giant optical activity in quasi-two-dimensional planar nanostructures \it Phys. Rev. Lett. \rm {\bf 95}(22) 227401
%
\bibitem{taka1} Takahashi S, Potts A, Bagnall D M, Zheludev N I and Zayats A V Nov 2005 Near-field polarization conversion in planar chiral nanostructures \it Opt. Comm. \rm {\bf 255}(1-3) 91--6
%
\bibitem{fedo_chiral} Fedotov V A, Mladyonov P L, Prosvirnin S L, Rogacheva A V, Chen Y and Zheludev N I Oct 2006 Asymmetric propagation of electromagnetic waves through a planar chiral structure \it Phys. Rev. Lett. \rm {\bf 97}(16) 167401
%
\bibitem{pott2} Potts A, Zhang W and Bagnall D M Apr 2008 Nonreciprocal diffraction through dielectric gratings with two-dimensional chirality \it Phys. Rev. \rm A \rm {\bf 77}(4) 043816
%
\bibitem{arnaNATO1996} Arnaut L R Mutual coupling in arrays of planar chiral structures. In: Priou A. {\it et al.} (eds) 1997 \it Advances in Complex Electromagnetic Materials \rm  Kluwer, Dordrecht, NL: NATO ASI Series III; High Technology {\bf 28} 293--309
%
\bibitem{arnaBIAN1997} Arnaut L R Multi-polarisability polyadics and spatial dispersion of bianisotropic and complex composite media \it Proc. ``Bianisotropics '97" Int. Conf. on Electromagnetics of Complex Media \rm (5--7 Jun. 1997, Glasgow, UK) 195--200
%
\bibitem{osip1} Osipov M A, Pickup B T, Fehervari M and Dunmur D A Jun 1998 Chirality measure and chiral order parameter for a two-dimensional system \it Molec. Phys. \rm {\bf 94}(2) 283--7
%
\bibitem{pott0} Potts A, Bagnall D M and Zheludev N I Feb 2004 A new model of geometrical chirality for two-dimensional continuous media and planar meta-materials \it J. Opt. {\rm A}: Pure Appl. Opt \rm {\bf 6}(2) 193--203
%
\bibitem{boru1} Boruhovich S P, Prosvirnin S L, Schwanecke A S and Zheludev N I Mar 2006 Multiplicative measure of planar chirality for 2D meta-materials \it Proc. Eur. Microw. Assoc. \rm {\bf 2}(1) 89--93
%
\bibitem{arnaACES1997} Arnaut L R Numerical multipole modelling of bianisotropic and complex materials \it Proc. 13th Annu. Rev. Progr. Appl. Comp. Electromagnetics (ACES'97) \rm (17--21 March 1997, Monterey, CA) 789--5
%
\bibitem{zhel_spaser} Zheludev N I, Prosvirnin S L, Papasimakis N and Fedotov V A Jun 2008 Lasing spaser \it Nature Photonics \rm {\bf 2} 351--4
%
\bibitem{wu1} Wu T T Nov 1962 Theory of thin circular loop antennas \it J. Math. Phys. \rm {\bf 3}(6) 1301--4
%
\bibitem{king1} King R W P and Harrison Ch W H 1969 \it Antennas and Waves \rm Cambridge, MA: MIT Press
%
\bibitem{lo1} Rispin L W and Chang D C Wire and loop antennas. In: Lo Y T and Lee S W (eds) 1988 \it Antenna Handbook \rm ch. 7. \rm New York, NY: Van Nostrand Reinhold
%
\bibitem{liis1} Jylh\"{a} L and Sihvola A Aug 2007 Equation for the effective permittivity of particle-filled composites for material design applications \it J. Phys. {\rm D}: Appl. Phys. \rm {\bf 40}(16) 4966--73
%
\bibitem{eyge1} Eyges L 1972 \it The Classical Electromagnetic Field \rm Reading, MA: Addison--Wesley
%
\bibitem{vanb2} van Bladel J 1991 \it Singular Electromagnetic Fields and Sources \rm Oxford, UK: Clarendon
%
\bibitem{arnaAEU98} Arnaut L R  Feb 1998 Recursive de-embedding procedure for computation of mutual coupling between bianisotropic or complex sources and scatterers \it AE\"{U} Int. J. Electron. Commun. \rm {\bf 52}(1) 1--8
%
\bibitem{arnaBIAN1998} Arnaut L R Analysis of an omnidirectional chiral structure \it Proc. ``Bianisotropics '98" Int. Conf. on Electromagnetics of Complex Media \rm (03--06 Jun. 1998, Braunschweig, Germany) 189--92
%
\bibitem{brew1} Brewitt-Taylor C R Modelling of helix-loaded chiral radar-absorbing layers. In: Priou A (ed) 1994 \it Progress in Electromagnetics Research \rm {\bf 9} Cambridge, MA: EMW Publishing, 289--310
%
\bibitem{park1} Parker E A and Hamdy S M A Aug 1981 Rings as elements for frequency selective surfaces \it Electron. Lett. \rm {\bf 17}(17) 612--14
%
\bibitem{schu1} Schuchinsky A G, Zelenchuk D E and Lerer A M Feb 2005 Enhanced transmission in microwave arrays of periodic sub-wavelength apertures \it J. Opt. {\rm A}: Pure Appl. Opt. \rm {\bf 7}(2) S102--9
%
\bibitem{munk1} Munk B A 2000 \it Frequency Selective Surfaces: Theory and Design \rm New York, NY: Wiley
%
\bibitem{blac1} Blackburn J F and Arnaut L R Oct 2005 Numerical convergence in periodic method of moments analysis of frequency-selective surfaces based on wire elements \it IEEE Trans. Antennas Propag. \rm {\bf 53}(10) 3308--15
%
\bibitem{maly1} Malyuskin O, Fusco V F and Schuchinsky A Nov-Dec 2008 Modelling of impedance-loaded wire frequency-selective surfaces with tunable reflection and transmission characteristics \it Int. J. Num. Modelling \rm {\bf 21}(6) 439--53
%
\bibitem{arnaPIER16} Arnaut L R and Davis L E Mutual coupling between bianisotropic particles: a theoretical study. \rm In: Kong J A (ed) 1997 \it Progress in Electromagnetics Research \rm {\bf 16} Cambridge, MA: EMW Publishing, 35--66
%
\bibitem{kont1} Kontorovich L V and Akilov G P 1964 \it Functional Analysis in Normed Spaces \rm Oxford, UK: Pergamon
%
\bibitem{buck1} Buckingham A D 1967 Permanent and induced molecular moments and long-range intermolecular forces \it Adv. Chem. Phys. \rm {\bf 12} 107--42
}
\end{thebibliography}
\end{document}